\documentclass[acmsmall]{acmart}

\AtBeginDocument{%
  \providecommand\BibTeX{{%
    \normalfont B\kern-0.5em{\scshape i\kern-0.25em b}\kern-0.8em\TeX}}}

\setcopyright{acmlicensed}
\acmDOI{10.1145/3643763}
\acmYear{2024}
\copyrightyear{2024}
\acmSubmissionID{fse24main-p653-p}
\acmJournal{PACMSE}
\acmVolume{1}
\acmNumber{FSE}
\acmArticle{37}
\acmMonth{7}
\received{2023-09-29}
\received[accepted]{2024-01-23}

\usepackage{lipsum}
\usepackage{balance}
\usepackage[normalem]{ulem}
\usepackage{multirow}
\usepackage{soul} 
\usepackage{color}
\usepackage{caption}
\usepackage{subcaption}
\usepackage{graphicx}
\usepackage{listings}
\usepackage{xcolor} 
\usepackage[most]{tcolorbox}
\usepackage{graphicx,wrapfig}
\usepackage{fontawesome5}
\definecolor{anti-flashwhite}{rgb}{0.95, 0.95, 0.96}
\definecolor{lightgray}{rgb}{0.83, 0.83, 0.83}
\definecolor{dkgreen}{rgb}{0,0.6,0}
\definecolor{gray}{rgb}{0.5,0.5,0.5}
\definecolor{mauve}{rgb}{0.58,0,0.82}
\definecolor{light-gray}{gray}{0.80}
\definecolor{gray}{rgb}{0.4,0.4,0.4}
\definecolor{darkblue}{rgb}{0.0,0.0,0.6}
\definecolor{cyan}{rgb}{0.0,0.6,0.6}
\definecolor{darkred}{rgb}{128.0,0.0,0.0}

\lstdefinestyle{mystyle}{
    frame=single,
    language=Java,
    aboveskip=3mm,
    belowskip=3mm,
    showstringspaces=false,
    escapeinside={(*@}{@*)},
    columns=flexible,
    basicstyle={\scriptsize \ttfamily},
    numbers=none,
    numberstyle=\tiny\color{black},
    commentstyle=\color{dkgreen},
    stringstyle=\color{mauve},
    breaklines=true,
    breakatwhitespace=true,
    tabsize=3,
    keywordstyle = {\color{blue}},
    keywordstyle = [2]{\color{blue}},
    keywordstyle = [3]{\color{mauve}},
    keywordstyle = [4]{\color{cyan}},
    otherkeywords = {requires, ensures, function, method, returns,exists, bool, string, then},
    morekeywords = [2]{decreases,forall, invariant, assert, print, expect}, 
    morekeywords = [3]{var, nat, Length}, 
    morekeywords = [4]{FindSmallest, ContainsZ,maxRunUppercase,Power,IsMinHeap, SplitAndAppend, FindMedian, PowerOfListElements, AppendArrayToSeq, ContainsSequence,InterleaveThreeSequences} 
}

\newcommand{\keyword}[1]{
    \textcolor{blue}{#1}
}
\newcommand{\literal}[1]{
    \textcolor{mauve}{#1}
}
\newcommand{\task}[1]{
    \textcolor{darkred}{#1}
}

\newcommand{\revise}[1]{
    \textcolor{black}{#1}
}

\begin{document}



\title{Towards AI-Assisted Synthesis of Verified Dafny Methods}



\author{Md Rakib Hossain Misu}
\authornote{\textbf{Corresponding author: Md Rakib Hossain Misu (mdrh@uci.edu)}}
\affiliation{
  \institution{University of California Irvine}
  \city{Irvine}
  \state{California}
  \country{USA}
}
\email{mdrh@uci.edu}
\orcid{0000-0002-7931-6782}

\author{Cristina V. Lopes}
\affiliation{
  \institution{University of California Irvine}
  \city{Irvine}
  \state{California}
  \country{USA}
}
\email{lopes@uci.edu}
\orcid{0000-0003-0551-3908}

\author{Iris Ma}
\affiliation{
  \institution{University of California Irvine}
  \city{Irvine}
  \state{California}
  \country{USA}
}
\email{huaiyaom@uci.edu}
\orcid{0009-0003-3699-7981}

\author{James Noble}
\additionalaffiliation{%
  \institution{Australian National University}
  \city{Canberra}
  \country{Australia}
}
\affiliation{%
  \institution{Creative Research \& Programming}
  \city{Wellington}
  \country{New Zealand}
}
\email{kjx@programming.ac.nz}
\orcid{0000-0001-9036-5692}



\renewcommand{\shortauthors}{M.R.H Misu, C.V.Lopes, I.Ma and J.Noble}

\newcommand{\gpt}{GPT-4~}
\newcommand{\palm}{PaLM-2~}

\begin{abstract}
  \noindent Large language models show great promise in many domains, including programming. A promise is easy to make but hard to keep, and language models often fail to keep their promises, generating erroneous code. A promising avenue to keep models honest is to incorporate formal verification: generating programs' specifications as well as code, so that the code can be proved correct with respect to the specifications. Unfortunately, existing large language models show a severe lack of proficiency in
verified programming. 
\par
In this paper, we demonstrate how to improve two pretrained models' proficiency in the Dafny verification-aware language. 
Using 178 problems from the MBPP dataset, we prompt two contemporary models (GPT-4 and PaLM-2) to synthesize Dafny methods. We use three different types of prompts: a direct \textit{Contextless} prompt; a \textit{Signature} prompt that includes a method signature and test cases, and a \textit{Chain of Thought (CoT)} prompt that decomposes the problem into steps and includes retrieval augmentation generated example problems and solutions. Our results show that GPT-4 performs better than PaLM-2 on these tasks, and that both models perform best with the retrieval augmentation generated CoT prompt. GPT-4 was able to generate verified,  human-evaluated, Dafny methods for 58\% of the problems, however GPT-4 managed only 19\% of the problems with the \textit{Contextless} prompt, and even fewer (10\%) for the \textit{Signature} prompt.  We are thus able to contribute 153 verified Dafny solutions to MBPP problems, 50 that we wrote manually and 103 synthesized by GPT-4.
\par
Our results demonstrate that the benefits of formal program verification are now within reach of code-generating large language models.
Likewise, program verification systems can benefit from large language models, whether to synthesize code wholesale, to generate specifications, or to act as a "programmer's verification apprentice", to construct annotations such as loop invariants which are hard for programmers to write or verification tools to find. Finally, we expect that the approach we have pioneered here --- generating candidate solutions that are subsequently formally checked for correctness ---  should transfer to other domains (e.g., legal arguments, transport signaling, structural engineering) where solutions must be correct, where that correctness must be demonstrated, explained and understood by designers and end-users.

\end{abstract}

\ccsdesc[500]{Software and its engineering~Formal software verification}

\keywords{Program Synthesis, Dafny, LLM, Program Verification}


\maketitle


\section{Introduction}
\label{sec:intro}
Formal verification is increasingly being adopted to support the development of high-quality, provably correct software. High-stakes domains, such as security-sensitive systems, cryptographic libraries, aerospace systems, and embedded software in medical devices, look to formal verification for correctness guarantees. 
\revise{
For instance, CompCert C~\cite{CompCert} is a formally verified C compiler that won the ACM Software Systems award in 2021\footnote{https://awards.acm.org/software-system}:
in a comprehensive study~\cite{DBLP:conf/pldi/YangCER11}, researchers found no bugs in the CompCert C compiler compared to the GCC~\cite{GCC} and LLVM~\cite{LLVM} tool-chain.  This study motivated Airbus to adopt CompCert C to help ensure safety and enhance aircraft performance \cite{airbusC2011}. The  seL4 project~\cite{DBLP:conf/sosp/KleinEHACDEEKNSTW09}, awarded the ACM Software System Award in 2022, resulted in a formally verified high-assurance, high-performance operating system microkernel employed to protect an autonomous helicopter against cyber-attacks~\cite{seL4}. In order to secure communication, both Chrome and Android use formally verified cryptographic code~\cite{DBLP:journals/sigops/ErbsenPGSC20}. Mozilla incorporated its verified cryptographic library for Firefox performance improvement~\cite{mozilla}.}
\par
Formal code verification consists of two parts that go hand-in-hand: formal \emph{specification} of software properties, and automated, or semi-automated, \emph{verification} of those properties. Over the past 50 years, verification has made a number of significant breakthroughs, first with the development of interactive theorem provers in the 1960s~\cite{Automath}, and then, at the turn of the millennium, with the development of Satisfiability Modulo Theory (SMT) solvers~\cite{SMT-Handbook2021}. Unfortunately writing program properties and proofs is still a creative, manual process that requires significant effort, experience, and expertise. Formal specification languages are closer to mathematics than to regular programming languages, and necessarily incorporate subtle concepts based on classical or constructive logics. For example, writing the proofs for the seL4 microkernel was an eleven person-year effort of a group of experts~\cite{DBLP:conf/sp/MurrayMBGBSLGK13}.
\revise{
The additional code required to verify the CompCert C compiler is more than three times the size of the compiler itself; and took three person-years to write,  approximately two lines of code per day~\cite{DBLP:journals/cacm/Leroy09}. That these systems have received important awards is a testament to the exceptional effort required.}
\par
In the universe of languages and systems for formal code verification, Dafny~\cite{dafnytsite} stands out as having made a significant effort towards usability by programmers. Dafny is a strongly typed imperative programming language with  functional and object-oriented features that supports code verification via Hoare Logic~\cite{DBLP:journals/cacm/Hoare69} assertions, preconditions, postconditions, and invariants --- often known as Design by Contract \cite{meyer1992applying}. Although algorithmic code in Dafny is similar to many other programming languages, writing the formal specifications and auxiliary verification assertions is still difficult~\cite{DBLP:conf/fsen/FariaA23,MPTP}. 
\par
Over the past year, Large Language Models (LLMs) have been revolutionizing both the software industry and research in software engineering. LLM-assisted tools such as GitHub's Copilot \cite{copilot} and Amazon CodeWhisper \cite{codewhisperer} have been applied to many software development tasks, including as code generation, transformation, summarization, documentation, code review, program repair, and synthesis. Although their performance is impressive for straightforward tasks in popular programming languages, given more complex problems or less familiar languages (such as Dafny), LLMs tend to produce "hallucinations"\footnote{Respectable software engineers call hallucinations "faults", 
"errors", or "bugs".} \cite{mckenna2023sources,truth-O-meter-PP23,hallucination2023,azaria2023internal} --- inventing control structures or APIs that do not exist in the target programming language, borrowing features from other languages, and often generating code that fails to type-check or even fails to parse.
\par
This paper describes our initial investigations into combining LLMs and program verification. Our main aim is to evaluate how well
LLMs can synthesize correct Dafny code, including the formal specifications and validation conditions so that the code passes the Dafny verifier: as part of that effort, we are also interested in how the additional logical rigor required by verification enhances or degrades generation performance. This paper makes the following contributions:
\vskip+3mm
\begin{tcolorbox}[width=\textwidth, colframe=black, colback=anti-flashwhite!30, boxsep=1mm, arc=1mm,bottom=0mm, left=3pt]
\begin{itemize}
    \item [\faUnlock*] We conduct the first empirical study of LLMs synthesizing verifiable Dafny methods.
    \item [\faUnlock*] We demonstrate that a prompt following the principles of \textit{Chain of Thought (CoT)} \cite{wei2022chain} with retrieval augmentation generated ~\cite{RAG} semantically similar  few shot examples explaining a problem decomposition step-by-step, can synthesize verified and correct Dafny methods with meaningful specifications for 58\% of problems in our test dataset.
    \item [\faUnlock*] We release the dataset \href{https://github.com/Mondego/dafny-synthesis}{\textbf{MBPP-DFY-153}}, a collection of \textbf{153} programming problems with specifications, solutions and tests in/for Dafny, based on the MBPP (Mostly Basic Python Programming) dataset curated by Google Research~\cite{DBLP:journals/corr/abs-2108-07732}. 
    \item [\faUnlock*] We exhibit that program verification techniques can help ensure that LLMs' synthesized code is correct according to the specifications.
\end{itemize}
\end{tcolorbox}
\vskip+2mm
\noindent Our conclusions lead to a number of actionable insights. Most obviously, our first contribution means that programmers developing Dafny systems could already consider using LLMs --- indeed, for simple cases, an LLM may be able to generate \textit{both the specifications of a Dafny method, and a verifiably correct  method implementation} that meets those specifications. For researchers in LLMs or verification, our second contribution, the dataset of \textbf{153} formally verified Dafny examples, should enable further development of prompts and models to improve the performance of LLMs.
\par
Ultimately, synthesis of verified methods (let alone whole programs) is a huge problem space consisting of multiple sub-problems: this paper is an initial exploration of this space.  Whereas our first two contributions speak to the end to end task of synthesizing specifications and implementations from a narrative prompt, our third contribution highlights that LLMs may prove more effective at tackling sub-parts of the problem: generating just specifications from narrative prompts, or code from specifications (with or without other prompts), generating parts of method bodies, or generating  only verification annotations, such as loop invariants or termination conditions.
\par
Taken together, these results lead to two intriguing speculative ideas: that efforts on LLMs for code should concentrate on generating verifiable programs; and that our approach of combining LLMs with program verification techniques should  transfer to other problem domains.

\section{Background \& Motivation}
\label{sec:background}
\subsection{Dafny: A Verification-Aware Programming Language}
Dafny~\cite{dafnytsite,dafny2023} is a verification-aware statically typed programming language which was first developed in Microsoft Research (MSR)~\cite{microsoftresearch}; it is currently being developed with the support of the Amazon Automated Reasoning research group~\cite{awsautomatedreasoning}. Dafny includes features drawn from programming paradigms including imperative, functional, and object-oriented programming. Dafny's tool-chain includes translators for generating executable code in several target languages including C\#, Java, JavaScript, Go and Python~\cite{dafny-github}. The distinguishing feature of Dafny is that it supports code verification via design by contract. For that, it follows the framework of Floyd-Hoare-style ~\cite{DBLP:journals/cacm/Hoare69} program verification with preconditions, postconditions, loop invariants and other high-level formal proof synthesis features such as pure functions, predicates, lemmas, and automated proof by induction.
\par
To develop a verified program, developers write Dafny code along with the specifications (pre-and post-conditions), and then add loop invariants and assertions that help the Dafny verifier prove the correctness of their code. While coding in VS Code, the Dafny static program verifier checks the functional correctness based on developers' defined specifications and annotations. Correctness means the absence of any runtime errors with respect to the formal pre- and postconditions, which means that the code does what the developers specify it to do. In order to confirm that methods' specifications hold, the Dafny program verifier first transforms the code  into an intermediate verification representation \cite{le2011boogie} that encodes the verification conditions in  predicate calculus~\cite{DBLP:journals/software/Leino17}, and then invokes the Z3 SMT solver~\cite{DBLP:conf/tacas/MouraB08} to prove the verification conditions.  The validity of these verification conditions implies the correctness of a program's code~\cite{DBLP:journals/software/Leino17}. Z3 cannot always prove the verification conditions, even where such a proof exists. In these situations, developers need to guide the solver, e.g.\ giving more context by writing auxiliary verification annotations such as assertions, functions, predicates, and lemmas. 
\\
\par
In recent years, Dafny has had major successes: Microsoft used Dafny to formally verify security libraries and kernels~\cite{DBLP:conf/sosp/KleinEHACDEEKNSTW09,DBLP:journals/sigops/ErbsenPGSC20}, distributed systems~\cite{DBLP:conf/sosp/HawblitzelHKLPR15}, and concurrent programs~\cite{DBLP:conf/osdi/HawblitzelHLNPZZ14}; Intel is developing its hardware encryption library using Dafny~\cite{DBLP:journals/iacr/YangWCCY23}; ConsenSys successfully applied Dafny for their Ethereum Virtual Machine (EVM) verification~\cite{DBLP:conf/fm/CassezFGPQ23}; Amazon implemented the Amazon Web Service (AWS) authorization and encryption logic in Dafny and deployed the Dafny-generated Java code into production~\cite{DBLP:conf/cav/Cook18,DBLP:conf/tacas/ChakarovFRR22}.

\subsection{Motivating Example}

To illustrate how a Dafny program is statically verified, Figure~\ref{fig:dafny-examples} shows a simple procedure to find, and return, the smallest number in a given input array. The method \texttt{FindSmallest} is formally

\begin{wrapfigure}{rb}{0.62\textwidth}
\input{code/examp_2_line_no}
\captionsetup{labelsep=colon, name=Figure}
\caption{An example Dafny method; \texttt{FindSmallest}}
\label{fig:dafny-examples}
\end{wrapfigure}
\setlength{\textfloatsep}{0pt}
%
%
%
%

specified not just by the types of input and output parameters, but also with  preconditions (with \texttt{requires} clause) and postconditions (with \texttt{ensures} clause). 

The postconditions are the properties we want to ensure are true when this method returns: what it means for this method to be ``correct.'' The Z3 SMT solver is not always able to infer postconditions, especially across imperative code. Figure~\ref{fig:dafny-examples}
contains a loop that modifies state: we must provide loop invariants that hold before and after every iteration of the loop. 
\par
The \texttt{Main} method is the entry point to the program; it invokes the method \texttt{FindSmallest} for some test inputs and checks (with assert) the outputs. Like other specifications in Dafny, \texttt{assert} statements are also statically checked and do not incur any runtime overhead. Writing Dafny code is similar to writing code in other imperative and object-oriented programming languages, however,  researchers observed that writing formal specifications and auxiliary verification annotations in Dafny is still difficult~\cite{DBLP:conf/fsen/FariaA23,MPTP}. We can see this in Figure~\ref{fig:dafny-examples}, which requires 6 lines of formal annotations to verify the 5 lines of the \texttt{FindSmallest} method body.

\subsection{Large Language Models (LLMs)}
\label{sec:llm}
Large Language Models (LLMs) are a class of deep neural networks trained on publicly available data to perform a variety of downstream tasks in natural language processing (NLP)  \cite{DBLP:conf/emnlp/MadaanZ0YN22}. In the past year, LLMs have significantly impacted a wide range of application domains including code related tasks such as code generation, comprehension, summarization, and transformation \cite{DBLP:conf/sigsoft/FirstRRB23,DBLP:conf/acl/2023/LiMaking,DBLP:conf/nips/WuJLRSJS22}. 
\par
The recent improvement of transformer based models~\cite{radford2018improving} such as GPT (Generative Pre-trained Transformer) advanced the use of LLMs in many coding related tasks compared to other pre-trained models~\cite{DBLP:journals/corr/abs-2302-12692,DBLP:journals/corr/abs-2303-08774}. For instance, CodeX~~\cite{DBLP:journals/corr/abs-2107-03374} is a descendant of GPT-3 trained on massive  publicly available code repositories in GitHub, and is capable of performing coding related tasks in various programming languages. Codex  powers GitHub Copilot~\cite{copilot}, an AI ``programmer's assistant'' designed to assist software developers.  OpenAI~\cite{openai} has now 
discontinued its Codex APIs, and 
suggested researchers adopt the successors of GPT-3 (GPT-3.5 and GPT-4).
Recently, Google released PaLM-2~\cite{DBLP:journals/corr/abs-2305-10403}, a successor to their earlier PaLM~\cite{DBLP:journals/corr/abs-2204-02311} model which is capable of code generation and reasoning tasks.
\par
Unfortunately, the applicability of LLMs to \textit{formally verified} programming languages  has not been studied extensively ~\cite{DBLP:conf/sigsoft/FirstRRB23}. Languages that support formal verification are not popular, so there is a paucity of training data available. LLMs have, however, been found to be capable of working as generic reasoning engines solving complex problems and proving theorems~\cite{DBLP:conf/acl/2023/LiMaking,DBLP:conf/nips/WuJLRSJS22}. These recent successes motivated us to explore how contemporary LLMs can contribute to formal software development in Dafny.

\subsection{Prompt Engineering}
Prompt engineering~\cite{bach2022,dang2022prompt,deckers2023,hou-etal-2022-metaprompting,jiang2022,liu2022},
coined by Gwern Branwen in 2020\footnote{https://gwern.net/gpt-3}, describes the incremental and iterative 
process of crafting inputs (``prompts'') for  LLMs.  LLMs commonly support conversational interaction:
we give them a prompt, and they reply with a related text or image, plausibly derived from the prompt. To apply LLMs to specific technical tasks, however, an informal, simplistic approach to prompting usually leads to  disappointing results. Effective prompts need to be structured to align with both the targeted task, and the LLM that will receive the prompt.
\par
Prompt engineering takes a systematic approach to creating prompts to convey tasks to pre-trained LLMs \cite{promptgramming2021,promptgramming2023}. To synthesize formally verified programs, prompts need to be engineered 
so that task descriptions are as unambiguous as possible~\cite{bach2022,dang2022prompt},
and solution examples must suit the target programming language (in our case Dafny).
Prompt engineering techniques can also train LLMs explicitly, e.g.\  through few-shot learning with a limited number of example problems and solutions: in our domain, textual problem description, method signature, and test cases.
\section{Study Design}
\label{sec:design}
This study aims to investigate the current capability and potentiality of LLMs at synthesizing formally verified methods in Dafny. Specifically, we seek to answer the following research questions.
\begin{itemize}
    \item \textbf{RQ1~[Contextless Prompting]:} \textit{How effective are LLMs at synthesizing formally verified Dafny methods from simple natural language problem descriptions?}
    \item \textbf{RQ2~[Signature Prompting]:} \textit{ How does additional context (method signature and tests) affect the synthesis outcome of LLMs?}
    \item \textbf{RQ3~[Dynamic Few-Shot Prompting]:} \textit{Do Chain of Thought (CoT) prompts supported by semantically similar few-shot examples and step-by-step problem decomposition improve the synthesis outcomes of LLMs?}
\end{itemize}

\subsection{Test Dataset: MBPP-san-DFY}
\revise{To conduct our study, we needed a collection of problems with natural language  descriptions and corresponding formally verified Dafny code. A search for Dafny code in GitHub yielded only 188 repositories, none of which had natural language specifications of the code. To use that code, we would have had to regenerate the informal natural language specifications from the formal code.}
\par
\revise{ Instead, we adopted the Mostly Basic Programming Problems (MBPP) benchmark suite~\cite{DBLP:journals/corr/abs-2108-07732}. MBPP is a collection of around 1,000 crowd-sourced algorithmic tasks with text description, function signature, concise implementation in Python, and three test cases. The MBPP tasks focus on basic arithmetic, array and integer sequences, and string processing. The textual descriptions of these tasks are designed to be concrete enough that an introductory Python programmer should be able to translate them into code, without any additional clarifications.
Solutions to 427 of the MBPP tasks have been inspected manually, and published as a subset called MBPP-sanitized (MBPP-san). We based our data on MBPP-san. Our goal is to synthesize \emph{Dafny} programs, rather than Python, so we removed 82  tasks that relied on Python-specific data types or external libraries\footnote{Dafny does not have a standard library, much less external libraries. Note that any Dafny libraries would  need to be formally-verified prior, or else the Dafny tasks would become much more complex than their Python versions.} such as \texttt{regex}, \texttt{math}, \texttt{heapq}, etc., leaving a dataset of 345 problems.}
\par
In order to keep our study's costs under budget, we furthered narrowed our focus
to a randomly selected subset of 228 of these 345 problems employing sampling with 99\% confidence interval and 5\% margin of error. Finally we randomly picked a further subset of 50 tasks 
to act as verified Dafny exemplars, leaving us with 178 remaining problems for testing. We manually translated each of those 178 problem descriptions, signatures, and test cases into Dafny, but we did not translate the solutions' Python code, or attempt to write specifications. This resulted in our main test dataset, MBPP-san-DFY: 178 instances of MBPP-san problems, each one with a description, method signature, and three test cases translated to suit Dafny.
\par
\revise{The MBPP problem set and Python solutions to the MBPP problems are publicly available, and it seems likely they would have been incorporated into the LLMs’ training sets. There are no Dafny implementations of MBPP, however, so our MBPP-san-DFY dataset should not increase data leakage within our experiment.}

\subsection{Human Written Dataset: MBPP-DFY-50}
\label{sec:MBPP_DFY_50}
To perform \textit{Dynamic Few-Shot} prompting, we needed a relatively large and varied collection of verified Dafny methods that could serve as exemplars to be included in prompts. A search for an existing suitable collection proved fruitless. Our best alternative was to translate a subset of MBPP-san into Dafny ourselves, but now also translating the Python solution for each problem into Dafny, then coming up with specifications (method pre- and postconditions, loop invariants) for each example, and finally doing whatever  annotation, correction, or rewriting was needed until the Dafny verifier would signal that the method was correct.
\par
In this process, we experienced first-hand how hard it is to formally specify postconditions and to provide loop invariants and hints so that correct code can pass the Dafny verifier. It took approximately 220 hours for the first two authors to create this dataset of 50 formally verified problems, even with the Python solutions, full access to Dafny documentation, books, blog posts by experts, and Stack Overflow.

\subsection{Pilot Study and LLM Selection}
Program synthesis in a verification-aware language requires knowledge of the target language syntax and semantics, plus the language implementation's  reasoning and theorem-proving abilities~\cite{MPTP,DBLP:conf/sigsoft/FirstRRB23}. To analyze LLMs' reasoning capability in Dafny, we first performed a pilot study using our handwritten dataset MBPP-DFY-50. We started by looking at GPT-4~\cite{DBLP:journals/corr/abs-2303-08774} and GPT-3.5~\cite{DBLP:journals/corr/abs-2107-03374} from OpenAI~\cite{openai}, PaLM-2~\cite{DBLP:journals/corr/abs-2305-10403} and PaLM~\cite{DBLP:journals/corr/abs-2204-02311} from Google, and CodeT5+~\cite{DBLP:conf/emnlp/0034WJH21}and CodeT5~\cite{DBLP:journals/corr/abs-2305-07922} from Salesforce, LLaMA and Code-LLaMA~\cite{roziere2023code} from Meta. All these LLMs have been explored in prior code generation, program synthesis, and reasoning-related studies.
\par
Our pilot study consisted of querying all these LLMs with the problem description, asking each LLM to solve each problem. We then analyzed the LLM's responses to get a sense of their reasoning abilities, and the quality of their approaches compared to our ground truth. \revise{
For these 50 problems, CodeT5, CodeT5+, LLaMA, Code-LLaMA, and PaLM were unable to synthesize a single syntactically correct Dafny method with specifications. Even GPT-3.5 could generate syntactically correct solutions for 5 of the 50 pilot study problems --- although GPT-4 and PaLM-2 could generate \textit{verified} Dafny methods in 16/50 and 7/50 problems respectively. Following these results, we decided to focus our investigation onto GPT-4 and PaLM-2 only.}

\begin{wrapfigure}{R}{0.60\textwidth}
\centering
\input{prompts/p1}
\input{prompts/p2}
\captionsetup{labelsep=colon, name=Figure}
\caption{Examples of \textit{Contextless} and \textit{Signature} prompts}
\label{fig:prompt_1_and_2}
\end{wrapfigure}
\setlength{\textfloatsep}{0pt}

\subsection{Prompt Design}
Based on the background literature and our pilot study, we knew that our experiment would depend  on the prompts we offered to the LLMs. We experimented with three different prompts:

\setlength{\textfloatsep}{0pt}

\subsubsection{\textbf{RQ1~[Contextless Prompt]:}}
To answer RQ1 we designed a simple \textit{Contextless}  prompt that identifies the programming language (Dafny), the problem description, and formatting instructions for the code response. Our prompt design is inspired by~\citet{DBLP:journals/corr/abs-2108-07732}, and works well for popular languages like Python, JavaScript, Java, C\#, and C++. The \textit{Contextless} prompt instructs LLMs to synthesize a Dafny method in a single step without any additional context. Figure~\ref{fig:prompt_1_and_2} (top) shows an example of a \textit{Contextless} prompt.  Writing implementation code in Dafny is similar to writing code in other high-level programming languages, so our goal with \textit{Contextless} prompt is to establish a baseline of the LLMs' performance in Dafny method synthesis.

\subsubsection{\textbf{RQ2~[Signature Prompt]:}}
For RQ2, we designed \textit{Signature} prompt with the required method signature and three test cases (see Figure~\ref{fig:prompt_1_and_2}, bottom) that extends \textit{Contextless} prompt. Code generation experiments for non-verified code often perform better when prompted with signature and test cases~\cite{DBLP:journals/corr/abs-2108-07732,DBLP:journals/corr/abs-2107-03374}; we hypothesized that adding method signature and test cases would also benefit LLMs generating code that will be formally verified.

\subsubsection{\textbf{RQ3~[Dynamic Few-Shot Prompt]:}}
\textit{Chain of Thought (CoT)} reasoning, which decomposes a problem into multiple intermediate steps, can lead to drastically improved performance for code synthesis tasks~\cite{DBLP:conf/emnlp/MadaanZ0YN22,wei2022chain,promptgramming2021,promptgramming2023}. Combined with relevant examples
(\textit{aka} few-shot prompting),
we can prompt LLMs to generate solutions to problems that have little coverage in the training data. Our \textit{Chain of Though (CoT)} prompts comprise five examples of tasks and solutions, where each example decomposes the problem 
step-by-step, and explains how to derive the implementation in Dafny.
The rest of this subsection describes in detail how we engineer CoT prompts:
Figure~\ref{fig:cot_prompt} (bottom) depicts one resulting CoT prompt
while Figure~\ref{fig:cot_prompt} (top) shows an intermediate specification prompt,
produced as one step in this process.
\par
The design of our CoT prompt was informed by our own experience with Dafny \cite{MPTP,LearnEm}. When we write a Dafny method, we start by defining the method's signature, followed by thinking about the pre- and postconditions that should hold. We then go on to  write the method body code, implementing algorithms   that should produce the expected results. Finally we address  validity, experimenting with assertions, loop invariants, etc., 
until the Dafny verifier is able to prove that the program is correct.

\paragraph{\textbf{Finding examples with Retrieval-Augmented Generation (RAG)}}
For a given problem, a prompt will be most effective when it includes semantically similar exemplars. While creating our MBPP-DFY-50 dataset, we observed that the postconditions are the most critical parts of Dafny method specifications, and we identified a pattern: methods having similar signatures and postconditions require similar Dafny solutions.
\par
\begin{figure}[t]
\includegraphics[width=\textwidth]{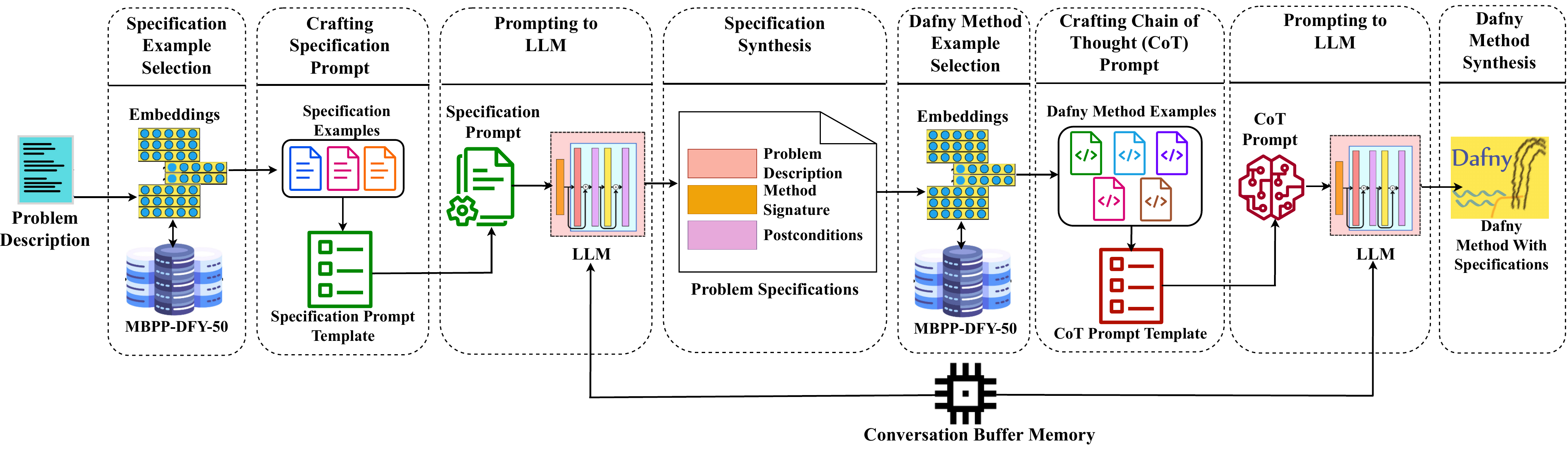}
\captionsetup{labelsep=colon, name=Figure}
\caption{A high-level architecture of the RAG prompt-chaining pipeline for Dynamic Few-Shot Prompting.}
\label{fig:pipeline_dynamic_few_shot}
\end{figure}

Given that we were already using LLMs to find solutions, we were also able to employ the same LLMs to find prompts, a technique known as prompt chaining~\cite{wei2022chain} or model cascading \cite{dohan2022language}.  For each new problem, in a first phase 
we prompt the LLM to generate method specifications (signature and postconditions).
Then in a second phase, we used the results of the first phase to prompt for Dafny method implementations. The inclusion of semantically-similar context in prompts is known as Retrieval Augmented Generation (RAG)~\cite{RAG}. Figure~\ref{fig:pipeline_dynamic_few_shot} demonstrates the resulting high-level architecture of our RAG pipeline for prompt chaining.

\par
For a given problem, in the first phase we prompt the LLM to generate the method specifications ---signature and postconditions. To teach LLM how to generate such specifications, we provide three examples of problem descriptions, method signatures, and postconditions, taken from our MBPP-DFY-50 dataset. We select these examples based on the semantic similarity of the problem descriptions. Figure~\ref{fig:cot_prompt} (top) shows an example of a first phase specification prompt. 

\setlength{\textfloatsep}{0pt}
\begin{figure}[]
    \centering
    \input{prompts/p3.1}
    \input{prompts/p3.2}
    \vskip-5mm
    \captionsetup{labelsep=colon, name=Figure}
    \caption{Examples of Specification Prompt (top) and CoT Prompt (bottom)}
    \label{fig:cot_prompt}
\end{figure}
\setlength{\textfloatsep}{0pt}%
In the second phase, we search for semantically similar Dafny methods in our MBPP-DFY-50 dataset, incorporating the LLM's first phase response. We use the OpenAI embedding API to compute semantic similarity based on cosine distance. For the  specification generated in the first phase, we select the top five Dafny similar  examples from our MBPP-DFY-50 dataset. We use five examples so that the whole prompt fits inside the LLMs' context length. With these examples, we then prepare a CoT prompt that first instructs the LLM to synthesize the specifications according the problem decomposition steps, and then synthesize a Dafny method that satisfies these specifications --- see Figure~\ref{fig:cot_prompt} (bottom).

\subsection{Evaluation Metrics}
We evaluate LLMs' synthesized methods using the Dafny verifier. The verifier reports whether Dafny code is (a) syntactically correct and (b) verified. We report the synthesis performance using the \textit{pass@k} metric~\cite{DBLP:conf/nips/KulalPC0PAL19,DBLP:journals/corr/abs-2107-03374, DBLP:journals/corr/abs-2303-08774, DBLP:journals/corr/abs-2305-07922}. In our study, we use \textit{verify@k} metric (similar to \textit{pass@k}), where \textit{k} sample methods are synthesized per problem; a problem is considered to be verified if any attempt verifies. 
\par
Automated quantitative metrics are only a weak proxy for performance of LLMs in this task. In one extreme, the synthesized Dafny code may not contain any formal specifications whatsoever; in those cases, the verifier will report verification success. Synthesized methods may contain only weak and uninteresting formal specifications that are trivial to verify. The most challenging part of this study is to check whether LLMs are able to generate the ``right'' and ``strong'' formal specifications --- in particular, postconditions. In order to assess the semantic correctness of the generated verification code, we manually reviewed all verified methods. We also manually inspected all the generated methods that failed to verify, to identify what types of errors prevented verification.
\subsection{Temperature Tuning}
\label{sec:temperature_tuning}
\begin{figure}[t]
\includegraphics[width=\linewidth]{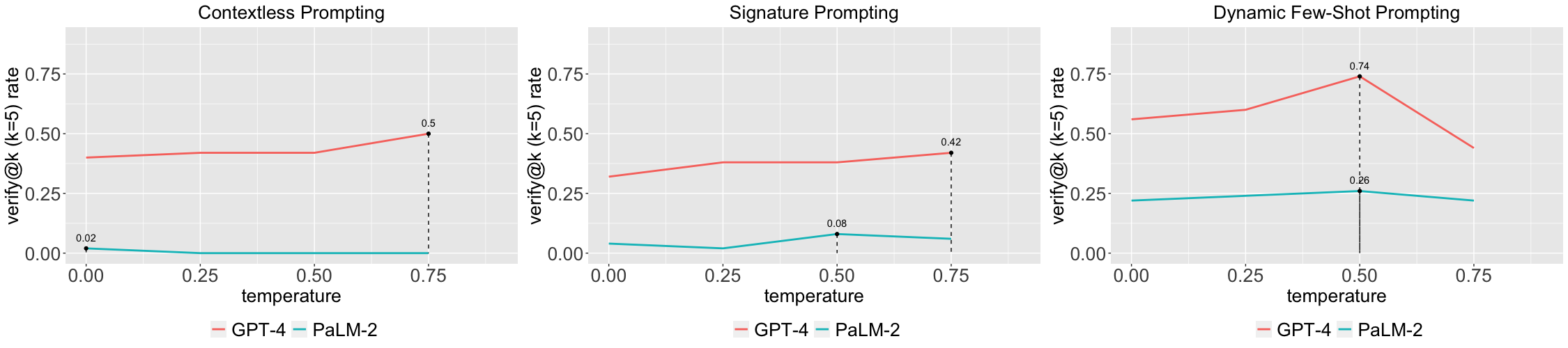}{}
\captionsetup{labelsep=colon, name=Figure}
\caption{Temperature tuning for \gpt and \palm on a sample of 50 problems at four different temperatures $T\in\{0.0,0.25,0.5,0.75\}$ for three different prompts. The vertical axis shows the verified Dafny method synthesis rate in verify@k (k=5).}
\label{fig:temperature-tuning}
\vspace{-2mm}
\end{figure}

The ``temperature'' is an important parameter of all LLMs that controls the randomness of the decoding process~\cite{DBLP:journals/corr/abs-2303-08774,DBLP:journals/corr/abs-2305-10403}. In doing experiments with LLMs, it is important to identify the best temperature for each experiment. Therefore, before running our experiments on the entire MBPP-san-DFY dataset, we first performed a temperature tuning experiment. We chose a random sample of 50 problems from our test dataset MBPP-san-DFY of 178 problems employing the sampling procedures with 90\% confidence interval and 10\% margin of error. Next, for each type of prompt, we executed these 50 problems at four different temperatures \textit{$T\in\{0.0, 0.25, 0.5, 0.75\}$} in both \gpt and \palm. We evaluated the synthesized Dafny methods with our evaluation metrics \textit{verify@k} where $k\in\{1,3,5\}$. \revise{Figure~\ref{fig:temperature-tuning}} shows that in \textit{Contextless} and \textit{Signature} prompts, \gpt performs best at \textit{T=0.75}, although with \textit{Dynamic Few-Shot} prompt, \gpt is most effective at \textit{T=0.5}. \palm demonstrates better performance with \textit{T=0.0} for \textit{Contextless} prompt, but in \textit{Signature} and \textit{Dynamic Few-Shot} prompts \palm works better at \textit{T=0.5}.

\setlength{\textfloatsep}{5pt}

\subsection{Experimental Setup}
We developed scripts to execute  prompts at different temperatures through OpenAI's API for GPT-4 and Google's API for PaLM-2. For verifying the code, we employed the latest Dafny 4.0.0 version \footnote{\href{https://dafny.org/blog/2023/03/03/dafny-4-released/}{https://dafny.org/blog/2023/03/03/dafny-4-released/}}. Dafny 4.0.0, released in March 2023, broke backwards compatibility, so it is likely neither \gpt nor \palm would have this latest version in their training data. In our prompts and few-shot examples we did not specify which version of Dafny should be used. For executing the experiments and verifying synthesized Dafny methods, we used a MacBook Pro (13-inch, 2020), having 2 GHz Quad-Core Intel® Core™ i5 processor, 16GB RAM, 512 GB SSD and running macOS Ventura 13.4.

\section{Results and Analysis}
\label{sec:results}
To evaluate LLM synthesis of  Dafny methods, we queried all 178 problems in three prompts with their tuned temperatures (as described in Section~\ref{sec:temperature_tuning}). We noticed that LLMs could not generate any Dafny methods if they were unable to generate the correct postconditions or if they had insufficient context data to solve those problems. In such cases the LLMs returned a textual explanation why they could not generate any Dafny code; we filtered out those responses. We then analyzed the verification results quantitatively and qualitatively.

\subsection{Quantitative Analysis}
In this section, we explain and compare our quantitative analysis across three different prompts for \gpt and \palm.
\label{sec:quantitative_analysis}
\subsubsection{\textbf{Verification:}}
Table~\ref{tab:verification_results} summarizes the overall verification results for three prompts queried with their tuned temperatures. For example, in \textit{Contextless} prompting at \textit{k=5}, \gpt~synthesized 104 verified methods out of 178 problems, achieving a verification success rate 58.42\%. We notice that the verification success rate improves with the increasing number of attempts. For instance, in \textit{Contextless} prompt at \textit{k=1}, \gpt's verification success rate is 32.58\% whereas at \textit{k=3} and \textit{k=5} it jumps to 53.37\% and 58.42\% respectively. \palm was unable to generate a single verified method for \textit{Contextless} prompting even with several attempts. In \textit{Signature} prompting, the verification success rate also improves with the number of attempts. Here, \gpt~generated 95 verified Dafny methods (53.37\%), which is slightly lower than \textit{Contextless} prompting. \palm~ gets a boost at \textit{k=5}, synthesizing 12~(6.74\%) verified methods. Both \gpt and \palm attain the best performance in \textit{Dynamic Few-shot} prompting. \gpt verified 114 methods (64.04\%) at \textit{k=5} that surpasses \textit{Contextless} prompting's performance. \palm achieved significant improvement with a 29.21\% verification success which is around four times of its  performance with \textit{Signature} prompting.
\begin{table}[t]
\setlength{\tabcolsep}{7pt}
\footnotesize
\captionsetup{labelsep=colon}
\caption{Summary of verified Dafny method synthesis on 178 problems. For each prompt, LLMs responses were generated with their tuned temperature (\textit{T}) at \textit{verify@k} (where $k \in \{1, 3, 5\}$)}
\label{tab:verification_results}
\begin{tabular}{|cclllcll|}
\hline
\multicolumn{1}{|c|}{\textbf{Prompts}}                           & \multicolumn{3}{c|}{\textbf{\gpt}}                                                                                      & \multicolumn{1}{l|}{\multirow{5}{*}{}} & \multicolumn{3}{c|}{\textbf{\palm}}                                                               \\ \cline{1-4} \cline{6-8} 
\multicolumn{1}{|c|}{\multirow{4}{*}{\textbf{Contextless}}}      & \multicolumn{1}{c|}{Temperature}             & \multicolumn{2}{c|}{\textit{verify@k}}                                    & \multicolumn{1}{l|}{}                  & \multicolumn{1}{l|}{Temperature}             & \multicolumn{2}{c|}{\textit{verify@k}}              \\ \cline{2-4} \cline{6-8} 
\multicolumn{1}{|c|}{}                                  & \multicolumn{1}{c|}{\multirow{3}{*}{\textit{T=0.75}}} & \multicolumn{1}{c|}{k=1} & \multicolumn{1}{l|}{58/178~(32.58\%)}  & \multicolumn{1}{l|}{}                  & \multicolumn{1}{c|}{\multirow{3}{*}{\textit{T=0.00}}} & \multicolumn{1}{c|}{k=1} & 0/178~(0.00\%)   \\ \cline{3-4} \cline{7-8} 
\multicolumn{1}{|c|}{}                                  & \multicolumn{1}{c|}{}                        & \multicolumn{1}{c|}{k=3} & \multicolumn{1}{l|}{92/178~(51.64\%)}  & \multicolumn{1}{l|}{}                  & \multicolumn{1}{c|}{}                        & \multicolumn{1}{c|}{k=3} & 0/178~(0.00\%)   \\ \cline{3-4} \cline{7-8} 
\multicolumn{1}{|c|}{}                                  & \multicolumn{1}{c|}{}                        & \multicolumn{1}{c|}{k=5} & \multicolumn{1}{l|}{104/178~(\textbf{58.42\%})} & \multicolumn{1}{l|}{}                  & \multicolumn{1}{c|}{}                        & \multicolumn{1}{c|}{k=5} & 0/178~(\textbf{0.00\%})   \\ \hline
\multicolumn{8}{|l|}{}                                                                                                                                                                                                                                                                                         \\ \hline
\multicolumn{1}{|c|}{\multirow{4}{*}{\textbf{Signature}}}        & \multicolumn{1}{l|}{Temperature}             & \multicolumn{2}{c|}{\textit{verify@k}}                                    & \multicolumn{1}{l|}{\multirow{4}{*}{}} & \multicolumn{1}{l|}{Temperature}             & \multicolumn{2}{c|}{\textit{verify@k}}              \\ \cline{2-4} \cline{6-8} 
\multicolumn{1}{|c|}{}                                  & \multicolumn{1}{c|}{\multirow{3}{*}{\textit{T=0.75}}} & \multicolumn{1}{l|}{k=1} & \multicolumn{1}{l|}{59/178~(33.14\%)}  & \multicolumn{1}{l|}{}                  & \multicolumn{1}{c|}{\multirow{3}{*}{\textit{T=0.50}}} & \multicolumn{1}{l|}{k=1} & 12/178~(6.74\%)  \\ \cline{3-4} \cline{7-8} 
\multicolumn{1}{|c|}{}                                  & \multicolumn{1}{c|}{}                        & \multicolumn{1}{l|}{k=3} & \multicolumn{1}{l|}{88/178~(49.43\%)}  & \multicolumn{1}{l|}{}                  & \multicolumn{1}{c|}{}                        & \multicolumn{1}{l|}{k=3} & 9/178~(5.05\%)   \\ \cline{3-4} \cline{7-8} 
\multicolumn{1}{|c|}{}                                  & \multicolumn{1}{c|}{}                        & \multicolumn{1}{l|}{k=5} & \multicolumn{1}{l|}{95/178~(\textbf{53.37\%})}  & \multicolumn{1}{l|}{}                  & \multicolumn{1}{c|}{}                        & \multicolumn{1}{l|}{k=5} & 12/178~(\textbf{6.74\%})  \\ \hline
\multicolumn{8}{|l|}{}                                                                                                                                                                                                                                                                                         \\ \hline
\multicolumn{1}{|l|}{\multirow{4}{*}{\textbf{Dynamic Few-Shot}}} & \multicolumn{1}{l|}{Temperature}             & \multicolumn{2}{c|}{\textit{verify@k}}                                    & \multicolumn{1}{l|}{\multirow{4}{*}{}} & \multicolumn{1}{l|}{Temperature}             & \multicolumn{2}{c|}{\textit{verify@k}}              \\ \cline{2-4} \cline{6-8} 
\multicolumn{1}{|l|}{}                                  & \multicolumn{1}{c|}{\multirow{3}{*}{\textit{T=0.50}}} & \multicolumn{1}{l|}{k=1} & \multicolumn{1}{l|}{86/178~(48.31\%)}  & \multicolumn{1}{l|}{}                  & \multicolumn{1}{c|}{\multirow{3}{*}{\textit{T=0.50}}} & \multicolumn{1}{l|}{k=1} & 30/178~(16.85\%) \\ \cline{3-4} \cline{7-8} 
\multicolumn{1}{|l|}{}                                  & \multicolumn{1}{c|}{}                        & \multicolumn{1}{l|}{k=3} & \multicolumn{1}{l|}{105/178~(58.98\%)} & \multicolumn{1}{l|}{}                  & \multicolumn{1}{c|}{}                        & \multicolumn{1}{l|}{k=3} & 43/178~(24.15\%) \\ \cline{3-4} \cline{7-8} 
\multicolumn{1}{|l|}{}                                  & \multicolumn{1}{c|}{}                        & \multicolumn{1}{l|}{k=5} & \multicolumn{1}{l|}{114/178~(\textbf{64.04\%})} & \multicolumn{1}{l|}{}                  & \multicolumn{1}{c|}{}                        & \multicolumn{1}{l|}{k=5} & 52/178~(\textbf{29.21\%}) \\ \hline
\end{tabular}
\end{table}

\subsubsection{\textbf{Postconditions:}}
In a naive interpretation, these results could indicate that the complicated RAG prompt is not worth it, at least for \gpt; however, that is not the case. Verification success does not mean that the solution was properly formulated. Postconditions are essential in formally specifying what methods are supposed to do. The absence of postconditions is a strong sign that the solution is not usefully verified: methods with no specifications trivially avoid being wrong.
\par
For both the verified and unverified methods, we looked into the formal specification synthesis results. In \textit{Contextless} prompting as shown in Table~\ref{tab:specification_result}, we noticed that, although \gpt~generated 104 verified methods, only 56 of them contain postconditions. Since the other 48 verified methods have no postconditions, they cannot be considered formally verified. Regarding unverified methods, \gpt generates postconditions for 65.75\% of them. Similar to \gpt, \palm also shows better effectiveness (59.09\%) in synthesizing postconditions in unverified methods. Since these methods are unverified, however, it may happen that either the postconditions are wrong, that the syntactic or semantic implementation of these methods are  wrong, or that both specifications and method implementations are correct, but that the Dafny verifier was unable to prove the method implementation in fact maintained the  postconditions. In our qualitative analyses we manually examine these methods and their specifications  to identify the actual reasons (see Section~\ref{sec:qualitative_analysis}).
\par
In \textit{Signature} prompting, the presence of postconditions gets worse. Only 30 out of 95 synthesis methods from \gpt have postconditions whereas for \palm only 1 method out of 12 contains postconditions. For unverified methods the presence of postconditions drastically drops for both \gpt (31.57\%) and \palm (38.27\%), indicating that any additional contexts in the prompt did not bring any substantial benefit.
\par
With \textit{Dynamic Few-Shot} RAG prompting, however, the specification synthesis results improve significantly for both \gpt and \palm. As shown in Table~\ref{tab:specification_result}, 100\% of both verified and unverified methods have postconditions. The primary reason for having postconditions in all synthesized methods is the few-shot examples with the CoT process. We designed this prompt to dynamically select five methods based on their postconditions' semantic similarity, and each of these examples includes a CoT process describing how to determine the postconditions in natural language, and then in Dafny in the implementation. The design of this prompt makes the LLMs follow the manner by which the examples are presented.

\subsubsection{\textbf{Preconditions:}} In a Dafny method, preconditions are only required when certain properties or assumptions must hold  before invoking the method.  For example, if a method reads an array element, preconditions can be used to help ensure that the method does not attempt to access array elements that are out-of-bounds.
\par
In our experiment with \textit{Contextless} prompting, we found that 61.53\% of \gpt generated verified methods contain preconditions and 39.42\% contain both pre-and postconditions (Table~\ref{tab:specification_result}). Like postconditions, the presence of preconditions drops in \textit{Signature} prompting where only 48.42\% of verified methods contain preconditions and 22.10\% have both pre-and postconditions. Unfortunately, in \textit{Contextless} and \textit{Signature} prompting, \palm specification synthesis performance is trivial. For both \gpt and \palm, however, \textit{Dynamic Few-Shot} RAG prompting enhances the specification synthesis performance. As shown in Table~\ref{tab:specification_result}, on  average 64\% of verified and unverified programs for both \gpt and \palm contain pre-and postconditions. Once again, \textit{Dynamic Few-Shot} prompts perform significantly better than \textit{Contextless} and \textit{Signature} prompts.

\begin{table}[t]
\footnotesize
\captionsetup{labelsep=colon}
\caption{Summary of specifications synthesis result, in both verified and unverified methods. The number in the denominators represents the total number of verified and unverified methods respectively. For each prompt, LLMs responses were generated with their tuned temperatures at verify@k (k=5).}
\label{tab:specification_result}
\begin{tabular}{|cllllll|}
\hline
\multicolumn{2}{|c|}{\textbf{Prompts}}                                                                     & \multicolumn{2}{c|}{\textbf{\gpt}}                                                  & \multicolumn{1}{l|}{\multirow{6}{*}{}} & \multicolumn{2}{c|}{\textbf{\palm}}                                           \\ \cline{1-4} \cline{6-7} 
\multicolumn{1}{|c|}{\multirow{5}{*}{\textbf{Contextless}}}      & \multicolumn{1}{c|}{\textbf{Specifications}}     & \multicolumn{1}{c|}{Verified}         & \multicolumn{1}{c|}{Unverified}     & \multicolumn{1}{l|}{}                  & \multicolumn{1}{c|}{Verified}       & \multicolumn{1}{c|}{Unverified} \\ \cline{2-4} \cline{6-7} 
\multicolumn{1}{|c|}{}                                  & \multicolumn{1}{l|}{has postconditions} & \multicolumn{1}{l|}{56/104~(\textbf{53.84\%})}  & \multicolumn{1}{l|}{48/73~(65.75\%)} & \multicolumn{1}{l|}{}                  & \multicolumn{1}{c|}{N/A}            & 104/176~(59.09\%)                \\ \cline{2-4} \cline{6-7} 
\multicolumn{1}{|c|}{}                                  & \multicolumn{1}{l|}{has preconditions}  & \multicolumn{1}{l|}{64/104~(61.53\%)}  & \multicolumn{1}{l|}{47/73~(64.38\%)} & \multicolumn{1}{l|}{}                  & \multicolumn{1}{c|}{N/A}            & 2/176~(1.13\%)                   \\ \cline{2-4} \cline{6-7} 
\multicolumn{1}{|c|}{}                                  & \multicolumn{1}{l|}{has both}           & \multicolumn{1}{l|}{41/104~(39.42\%)}  & \multicolumn{1}{l|}{32/73~(43.83\%)} & \multicolumn{1}{l|}{}                  & \multicolumn{1}{c|}{N/A}            & 2/176~(1.13\%)                   \\ \cline{2-4} \cline{6-7} 
\multicolumn{1}{|c|}{}                                  & \multicolumn{1}{l|}{has invariants}     & \multicolumn{1}{l|}{40/104~(38.46\%)}  & \multicolumn{1}{l|}{28/73~(38.35\%)} & \multicolumn{1}{l|}{}                  & \multicolumn{1}{c|}{N/A}            & 4/176~(2.27\%)                   \\ \hline
\multicolumn{7}{|l|}{}                                                                                                                                                                                                                                                                           \\ \hline
\multicolumn{1}{|c|}{\multirow{5}{*}{\textbf{Signature}}}        & \multicolumn{1}{c|}{\textbf{Specifications}}     & \multicolumn{1}{c|}{Verified}         & \multicolumn{1}{c|}{Unverified}     & \multicolumn{1}{l|}{\multirow{5}{*}{}} & \multicolumn{1}{c|}{Verified}       & \multicolumn{1}{c|}{Unverified} \\ \cline{2-4} \cline{6-7} 
\multicolumn{1}{|c|}{}                                  & \multicolumn{1}{l|}{has postconditions} & \multicolumn{1}{l|}{30/95~(\textbf{31.57\%})}   & \multicolumn{1}{l|}{22/80~(27.5\%)}  & \multicolumn{1}{l|}{}                  & \multicolumn{1}{l|}{1/12~(8.33\%)}   & 62/162~(38.27\%)                 \\ \cline{2-4} \cline{6-7} 
\multicolumn{1}{|c|}{}                                  & \multicolumn{1}{l|}{has preconditions}  & \multicolumn{1}{l|}{46/95~(48.42\%)}   & \multicolumn{1}{l|}{31/80~(38.75\%)} & \multicolumn{1}{l|}{}                  & \multicolumn{1}{l|}{1/12~(8.33\%)}   & 61/162~(37.65\%)                 \\ \cline{2-4} \cline{6-7} 
\multicolumn{1}{|c|}{}                                  & \multicolumn{1}{l|}{has both}           & \multicolumn{1}{l|}{21/95~(22.10\%)}   & \multicolumn{1}{l|}{19/80~(23.75\%)} & \multicolumn{1}{l|}{}                  & \multicolumn{1}{l|}{1/12~(8.33\%)}   & 61/162~(37.65\%)                 \\ \cline{2-4} \cline{6-7} 
\multicolumn{1}{|c|}{}                                  & \multicolumn{1}{l|}{has invariants}     & \multicolumn{1}{l|}{38/95~(40.00\%)}   & \multicolumn{1}{l|}{19/80~(23.75\%)} & \multicolumn{1}{l|}{}                  & \multicolumn{1}{l|}{0/12~(0.00\%)}   & 6/162~(3.70\%)                   \\ \hline
\multicolumn{7}{|l|}{}                                                                                                                                                                                                                                                                           \\ \hline
\multicolumn{1}{|c|}{\multirow{5}{*}{\textbf{Dynamic Few-Shot}}} & \multicolumn{1}{c|}{\textbf{Specifications}}     & \multicolumn{1}{c|}{Verified}         & \multicolumn{1}{c|}{Unverified}     & \multicolumn{1}{l|}{\multirow{5}{*}{}} & \multicolumn{1}{c|}{Verified}       & \multicolumn{1}{c|}{Unverified} \\ \cline{2-4} \cline{6-7} 
\multicolumn{1}{|c|}{}                                  & \multicolumn{1}{l|}{has postconditions} & \multicolumn{1}{l|}{114/114~(\textbf{100\%})} & \multicolumn{1}{l|}{61/61~(\textbf{100\%})} & \multicolumn{1}{l|}{}                  & \multicolumn{1}{l|}{52/52~(\textbf{100\%})} & 123/123~(\textbf{100\%})                \\ \cline{2-4} \cline{6-7} 
\multicolumn{1}{|c|}{}                                  & \multicolumn{1}{l|}{has preconditions} & \multicolumn{1}{l|}{74/114~(64.91\%)}  & \multicolumn{1}{l|}{42/61~(68.85\%)} & \multicolumn{1}{l|}{}                  & \multicolumn{1}{l|}{35/52~(67.30\%)} & 99/123~(80.48\%)                 \\ \cline{2-4} \cline{6-7} 
\multicolumn{1}{|c|}{}                                  & \multicolumn{1}{l|}{has both}           & \multicolumn{1}{l|}{74/114~(64.91\%)}  & \multicolumn{1}{l|}{42/61~(68.85\%)} & \multicolumn{1}{l|}{}                  & \multicolumn{1}{l|}{35/52~(67.30\%)} & 99/123~(80.48\%)                 \\ \cline{2-4} \cline{6-7} 
\multicolumn{1}{|c|}{}                                  & \multicolumn{1}{l|}{has invariants}     & \multicolumn{1}{l|}{57/114~(50.0\%)}   & \multicolumn{1}{l|}{57/61~(93.44\%)} & \multicolumn{1}{l|}{}                  & \multicolumn{1}{l|}{13/52~(25.0\%)}  & 90/123~(73.17\%)                 \\ \hline
\end{tabular}
\end{table}

\subsubsection{\textbf{Invariants:}} When a Dafny method has a loop, it should contain invariants that need to be maintained in each iteration of the loop.  As shown in Table~\ref{tab:specification_result}, in \textit{Contextless} prompting, an average 38\% \gpt synthesized methods contain invariants. In \textit{Signature} prompting, the number increases slightly for verified methods, however, in \textit{Dynamic Few-Shot} prompting, \gpt does far better at generating invariants. As shown in Table~\ref{tab:specification_result}, 50\% of verified and 93.44\% of unverified methods contain at least one loop invariant. On the other hand, \palm struggles to synthesize loop invariants for both verified and unverified methods in \textit{Contextless} and \textit{Signature} prompting. Again, \textit{Dynamic Few-Shot} prompting  improves its performance to have loop invariants in 25\% of verified and 73.17\% for unverified methods respectively.

\subsection{Qualitative Analysis}
\label{sec:qualitative_analysis}
This section describes our manual investigation of LLMs' Dafny synthesis performance. The first two authors manually examined all prompt responses, assessing the quality of the formal specifications, specifically the postconditions of the verified Dafny methods -- again, postconditions are the key element for a correct formal \emph{specification}. During this manual inspection, when we have any doubt about the implementation or specifications, we looked into our ground-truth dataset, MBPP-san, and discussed with each other. The two authors independently performed this manual tagging and reached a high Inter-Rater Reliability (IRR)~\cite{hallgren2012computing}(93.91\% Cohen Kappa~\cite{viera2005understanding}), which ended up with consensus at the end. We summarize our qualitative analysis findings in Table~\ref{tab:qualatative_result}. The complete analysis results are available in our companion repository~(Section~\ref{sec:data-availability}), including all prompts, LLMs' responses, verification results, and Dafny method snippets with our manual tagging.

\subsubsection{\textbf{Postconditions:}}
The goal of a postcondition is to make sure that the implementation is \textit{correct} based on the proofs. Here the term "\textit{correct}" implies that the postcondition is strong enough to precisely capture all possible output values. \revise{A postcondition can be weak or less restrictive that allows a boarder range of possible outcomes. In some cases, a postcondition might be incorrect to accurately capture the intended outcomes and making it difficult or impossible to prove.}
\par
We tagged the quality of postconditions into three groups, \texttt{Strong}, \texttt{Weak} and \texttt{Wrong}. In \textit{Signature} prompting, \palm had only one verified method and that has \texttt{Strong} postconditions. With regard to, \textit{Dynamic Few-Shot} prompt, we found that in 14 cases, \palm misunderstood problems' description and generated wrong method implementation for those problems. However, it generated \texttt{Strong} postconditions in 35 methods and \texttt{Weak} postconditions in 3 methods only. Although \gpt got 56 verified programs in \textit{Contextless} prompting, we encountered 15 of them have \texttt{Weak} postconditions. Table~\ref{tab:dfy_example_snippets} exhibits such an example of \texttt{Weak} postconditions. On the other hand in \textit{Signature} prompting 9 of verified methods contain \texttt{Weak} postconditions which are very trivial specifications to verify correct implementation.
\par
In \textit{Dynamic Few-Shot}, \gpt's specification synthesis result is impressive. Only 3 of verified methods had \texttt{Weak} postconditions and  in 8 cases it misunderstood the problems' description, generating wrong methods. One interesting example of such cases is illustrated in Table~\ref{tab:dfy_example_snippets}. Specifically in \textit{Dynamic Few-Shot} RAG prompt, 90.35\% (103/114) of verified methods synthesized from \gpt contain \texttt{Strong} postconditions. \revise{Considering the overall dataset, 58\% of methods (103/178) had strong (i.e. correct) specifications that successfully verified.}

\begin{table}[t]
\setlength{\tabcolsep}{4pt}
\footnotesize
\captionsetup{labelsep=colon}
\caption{Summary of qualitative analysis of specifications synthesis in verified methods. The number in the denominator represents the total number of verified methods containing specifications. For each prompt, LLM responses were generated with their tuned temperatures at verify@k (k=5).}
\label{tab:qualatative_result}

\begin{tabular}{|lllllllll|}
\hline
\multicolumn{1}{|l|}{\multirow{3}{*}{\textbf{Specifications}}} & \multicolumn{1}{l|}{\multirow{3}{*}{\textbf{Groups}}} & \multicolumn{3}{c|}{\textbf{GPT-4}}                                                                                                                 & \multicolumn{1}{l|}{\multirow{6}{*}{}} & \multicolumn{3}{c|}{\textbf{PaLM-2}}                                                                                           \\ \cline{3-5} \cline{7-9} 
\multicolumn{1}{|l|}{}                                         & \multicolumn{1}{l|}{}                                 & \multicolumn{3}{c|}{Prompts}                                                                                                                        & \multicolumn{1}{l|}{}                  & \multicolumn{3}{c|}{Prompts}                                                                                                   \\ \cline{3-5} \cline{7-9} 
\multicolumn{1}{|l|}{}                                         & \multicolumn{1}{l|}{}                                 & \multicolumn{1}{l|}{Contextless} & \multicolumn{1}{l|}{Signature} & \multicolumn{1}{l|}{\begin{tabular}[c]{@{}l@{}}Dynamic\\ Few-Shot\end{tabular}} & \multicolumn{1}{l|}{}                  & \multicolumn{1}{l|}{Contextless} & \multicolumn{1}{l|}{Signature} & \begin{tabular}[c]{@{}l@{}}Dynamic\\ Few-Shot\end{tabular} \\ \cline{1-5} \cline{7-9} 
\multicolumn{1}{|l|}{\multirow{3}{*}{\textbf{Postconditios}}}  & \multicolumn{1}{l|}{\texttt{Strong}}                           & \multicolumn{1}{l|}{34/56}       & \multicolumn{1}{l|}{18/30}     & \multicolumn{1}{l|}{\textbf{103/114}~(\textbf{90.35\%})}                                                    & \multicolumn{1}{l|}{}                  & \multicolumn{1}{l|}{N/A}         & \multicolumn{1}{l|}{1/1}       & 35/52                                                      \\ \cline{2-5} \cline{7-9} 
\multicolumn{1}{|l|}{}                                         & \multicolumn{1}{l|}{\texttt{Weak}}                             & \multicolumn{1}{l|}{15/56}       & \multicolumn{1}{l|}{9/30}      & \multicolumn{1}{l|}{3/114}                                                      & \multicolumn{1}{l|}{}                  & \multicolumn{1}{l|}{N/A}         & \multicolumn{1}{l|}{0/1}       & 3/52                                                       \\ \cline{2-5} \cline{7-9} 
\multicolumn{1}{|l|}{}                                         & \multicolumn{1}{l|}{\texttt{Wrong}}                            & \multicolumn{1}{l|}{7/56}        & \multicolumn{1}{l|}{3/30}      & \multicolumn{1}{l|}{8/114}                                                      & \multicolumn{1}{l|}{}                  & \multicolumn{1}{l|}{N/A}         & \multicolumn{1}{l|}{0/1}       & 14/52                                                      \\ \hline
\multicolumn{9}{|l|}{}                                                                                                                                                                                                                                                                                                                                                                                                                                 \\ \hline
\multicolumn{1}{|l|}{\multirow{2}{*}{\textbf{Preconditions}}}  & \multicolumn{1}{l|}{\texttt{Required}}                         & \multicolumn{1}{l|}{19/41}       & \multicolumn{1}{l|}{10/21}     & \multicolumn{1}{l|}{\textbf{36/74~(48.64\%)}}                                                      & \multicolumn{1}{l|}{\multirow{2}{*}{}} & \multicolumn{1}{l|}{N/A}         & \multicolumn{1}{l|}{0/1}       & 6/35                                                       \\ \cline{2-5} \cline{7-9} 
\multicolumn{1}{|l|}{}                                         & \multicolumn{1}{l|}{\texttt{Optional}}                         & \multicolumn{1}{l|}{22/41}       & \multicolumn{1}{l|}{11/21}     & \multicolumn{1}{l|}{38/74}                                                      & \multicolumn{1}{l|}{}                  & \multicolumn{1}{l|}{N/A}         & \multicolumn{1}{l|}{1/1}       & 29/35                                                      \\ \hline
\multicolumn{9}{|l|}{}                                                                                                                                                                                                                                                                                                                                                                                                                                 \\ \hline
\multicolumn{1}{|l|}{\multirow{2}{*}{\textbf{Invariants}}}     & \multicolumn{1}{l|}{\texttt{Strong}}                           & \multicolumn{1}{l|}{10/27}       & \multicolumn{1}{l|}{5/16}      & \multicolumn{1}{l|}{\textbf{51/57~(89.47\%)}}                                                      & \multicolumn{1}{l|}{\multirow{2}{*}{}} & \multicolumn{1}{l|}{N/A}         & \multicolumn{1}{l|}{N/A}       & 13/13                                                      \\ \cline{2-5} \cline{7-9} 
\multicolumn{1}{|l|}{}                                         & \multicolumn{1}{l|}{\texttt{Weak}/\texttt{Wrong}}                       & \multicolumn{1}{l|}{17/27}       & \multicolumn{1}{l|}{11/16}     & \multicolumn{1}{l|}{6/57}                                                       & \multicolumn{1}{l|}{}                  & \multicolumn{1}{l|}{N/A}         & \multicolumn{1}{l|}{N/A}       & 0/13                                                       \\ \hline
\end{tabular}
\end{table}
\begin{table}[t]
\centering
\footnotesize
\setlength{\tabcolsep}{2pt}
\captionsetup{labelsep=colon}
\caption{Examples of LLMs synthesis \texttt{Strong}, \texttt{Weak} and \texttt{Wrong} postconditions; \texttt{Required} and \texttt{Optional} preconditions; invariants subsuming postconditions, and auxiliary definitions.}
\label{tab:dfy_example_snippets}
\begin{tabular}{|lll|}
\hline
\multicolumn{2}{|c|}{\textbf{Specifications}}                                                                                                                                                          & \multicolumn{1}{c|}{\textbf{LLM Synthesis Specifications in Dafny}} \\ \hline
\multicolumn{1}{|l|}{\multirow{3}{*}{\textbf{\begin{tabular}[c]{@{}l@{}}Post-\\ conditions\end{tabular}}}} & \multicolumn{1}{l|}{\texttt{Strong}}                                                               & \begin{lstlisting}[style = mystyle, frame=none]
//Write a (*@\textcolor{dkgreen}{method}@*) in Dafny to check if the given array represents min heap or not.
method IsMinHeap(a: array<int>) returns (result: bool)
    ensures result ==> forall i :: 0 <= i < a.Length / 2 ==> a[i] <= a[2*i + 1] 
                                && (2*i + 2 == a.Length || a[i] <= a[2*i + 2])
    ensures !result ==> exists i :: 0 <= i < a.Length / 2 && (a[i] > a[2*i + 1] 
                                || (2*i + 2 != a.Length && a[i] > a[2*i + 2]))
\end{lstlisting}                  \\ \cline{2-3} 
\multicolumn{1}{|l|}{}                                                                                     & \multicolumn{1}{l|}{\texttt{Weak}}                                                                 & \begin{lstlisting}[style = mystyle, frame=none]
/*Write a (*@\textcolor{dkgreen}{method}@*) in Dafny to interleave 3 sequence of the same length into 
a single flat sequence.*/
method InterleaveThreeSequences(seq1: seq<int>, seq2: seq<int>, seq3: seq<int>) 
    returns (interleavedSeq: seq<int>) 
    requires |seq1| == |seq2| == |seq3| 
    ensures |interleavedSeq| == 3 * |seq1|
//strong postcondition will also check the relative order of the  sequence values.
\end{lstlisting}                    \\ \cline{2-3} 
\multicolumn{1}{|l|}{}                                                                                     & \multicolumn{1}{l|}{\texttt{Wrong}}                                                                & \begin{lstlisting}[style = mystyle, frame=none]
//Write a (*@\textcolor{dkgreen}{method}@*) in Dafny to find the median of two sorted lists of same size.
method FindMedian(a: array<int>, b: array<int>) returns (median: int)
    //...MISUNDERSTANDING
    ensures median == if (a.Length % 2 == 0) 
                        then (a[a.Length / 2 - 1] + b[0]) / 2 else a[a.Length / 2]
\end{lstlisting}                   \\ \hline
\multicolumn{3}{|l|}{}                                                                                                                                                                                                                                    \\ \hline
\multicolumn{1}{|l|}{\multirow{2}{*}{\textbf{\begin{tabular}[c]{@{}l@{}}Pre-\\ conditions\end{tabular}}}}  & \multicolumn{1}{l|}{\texttt{Required}}                                                             &  \begin{lstlisting}[style = mystyle, frame=none]
//Write a (*@\textcolor{dkgreen}{method}@*) in Dafny to compute the n-th power of each number in a list.
function Power(base: int, exponent: int): int //.. Auxiliary Definitions
    requires exponent >= 0 
    { if exponent == 0 then 1 else base * Power(base, exponent-1) }
method PowerOfListElements(l: seq<int>, n: int) returns (result: seq<int>)
    requires n >= 0
    ensures |result| == |l|
    ensures forall i :: 0 <= i < |l| ==> result[i] == Power(l[i], n)
\end{lstlisting}
                                                \\ \cline{2-3} 
\multicolumn{1}{|l|}{}                                                                                     & \multicolumn{1}{l|}{\texttt{Optional}}                                                             &   \begin{lstlisting}[style = mystyle, frame=none]
//Write a (*@\textcolor{dkgreen}{method}@*) in Dafny to append the given array to the given sequence.
method AppendArrayToSeq(s: seq<int>, a: array<int>) returns (r: seq<int>)
    requires a != null
    /*In Dafny arrays and sequences are never null, 
    unless the (*@\textcolor{dkgreen}{method}@*) declares them as array? and seq?*/
\end{lstlisting}
                                        \\ \hline
\multicolumn{3}{|l|}{}                                                                                                                                                                                                                                    \\ \hline
\multicolumn{1}{|l|}{\textbf{Invariants}}                                                                  & \multicolumn{1}{l|}{\begin{tabular}[c]{@{}l@{}}\texttt{Subsume}\\ \texttt{Post-}\\ \texttt{conditions}\end{tabular}} &     \begin{lstlisting}[style = mystyle, frame=none]
/* Write a (*@\textcolor{dkgreen}{method}@*) in Dafny to check whether a list of sequence contains 
the given sequence or not.*/
method ContainsSequence(list: seq<seq<int>>, sub: seq<int>) returns (result: bool)
    ensures result <==> (exists i :: 0 <= i < |list| && sub == list[i])
{
    result := false;
    for i := 0 to |list|
        invariant 0 <= i <= |list|
        invariant result <==> (exists k :: 0 <= k < i && sub == list[k])
\end{lstlisting}                                             \\ \hline
\end{tabular}
\end{table}

\subsubsection{\textbf{Preconditions:}}
A precondition is an annotation that must be true before a Dafny method is invoked. When a method is called, it is caller's responsibility to make sure that the preconditions are satisfied. The caller of the method then gets to assume that the postconditions hold after the method returns~~\cite{dafny-github}. To keep in mind that, in our analysis, we did not call LLMs synthesis methods. In our manual investigation, we only tagged preconditions into two groups, (i) \texttt{Required}, which is necessary to prove the postconditions and implementation; (ii) \texttt{Optional}, its presence is not required for verification. Table~\ref{tab:qualatative_result} demonstrates the summary of our preconditions analysis. Examples of LLMs synthesis \texttt{Required} and \texttt{Optional} preconditions are illustrated in Table~\ref{tab:dfy_example_snippets}.

\subsubsection{\textbf{Invariants:}}
We reviewed loop invariants in all verified methods that have postconditions. To determine loop invariants in Dafny, a common trick is to work backwards from the postconditions and identify the loop guard~\cite{dafny-github}. While examining  responses, we recognized the same patterns followed by the LLMs. We observed that \gpt and \palm both determined loop invariants mostly based on the postconditions and in most of the problems loops invariants subsume some or whole part of the method's postconditions as shown in Table~\ref{tab:dfy_example_snippets}. We found that all synthesized verified methods in three different prompts containing \texttt{Strong} postconditions, also have correct loop invariants. And when the postconditions were \texttt{Weak} or \texttt{Wrong}, loop invariants followed the postconditions. In many cases we also noticed that \gpt guaranteed loop termination with the explicit usages of \texttt{decreases} statement on loop guards.

\subsubsection{\textbf{Auxiliary Definitions:}} Dafny, supports \texttt{predicate} and \texttt{function} to write modularized and readable  code~\cite{dafny-github}. When we reviewed all verified synthesized methods,  auxiliary definitions were only generated in response to \textit{Dynamic Few-Shot} prompts. We found that only 4 out of 52 \palm synthesized verified methods had auxiliary definitions whereas 20 out of 103 \gpt verified methods have the usages of \texttt{predicate} or \texttt{function}. Table~\ref{tab:dfy_example_snippets} shows an example of such auxiliary definitions with the usages of \texttt{function}.

\subsection{Error Analysis}
\begin{table}[t]
\setlength{\tabcolsep}{1.8pt}
\footnotesize
\captionsetup{labelsep=colon}
\caption{Summary of different types of errors found in unverified methods. The number in the denominators represents the total number of unverified methods. For each prompt, LLMs responses were generated with their tuned temperatures at verify@k (k=5).}
\label{tab:error_results_count}
\begin{tabular}{|lllllllll|}
\hline
\multicolumn{1}{|c|}{\multirow{3}{*}{\textbf{\begin{tabular}[c]{@{}c@{}}Error\\ Types\end{tabular}}}}         & \multicolumn{1}{c|}{\multirow{3}{*}{\textbf{Error Name}}} & \multicolumn{3}{c|}{\textbf{GPT-4}}                                                                                                                         & \multicolumn{1}{l|}{\multirow{5}{*}{}} & \multicolumn{3}{c|}{\textbf{PaLM-2}}                                                                                                       \\ \cline{3-5} \cline{7-9} 
\multicolumn{1}{|c|}{}                                                                                        & \multicolumn{1}{c|}{}                                     & \multicolumn{3}{c|}{Prompts}                                                                                                                                & \multicolumn{1}{l|}{}                  & \multicolumn{3}{c|}{Prompts}                                                                                                               \\ \cline{3-5} \cline{7-9} 
\multicolumn{1}{|c|}{}                                                                                        & \multicolumn{1}{c|}{}                                     & \multicolumn{1}{l|}{Contextless}    & \multicolumn{1}{l|}{Signature}      & \multicolumn{1}{l|}{\begin{tabular}[c]{@{}l@{}}Dynamic\\ Few-Shot\end{tabular}} & \multicolumn{1}{l|}{}                  & \multicolumn{1}{l|}{Contextless}      & \multicolumn{1}{l|}{Signature}        & \begin{tabular}[c]{@{}l@{}}Dynamic\\ Few-Shot\end{tabular} \\ \cline{1-5} \cline{7-9} 
\multicolumn{1}{|l|}{\multirow{2}{*}{\textbf{\begin{tabular}[c]{@{}l@{}}Compilation\\ Errors\end{tabular}}}}  & \multicolumn{1}{l|}{\texttt{parse errors}}                         & \multicolumn{1}{l|}{\textbf{51/73}} & \multicolumn{1}{l|}{\textbf{44/80}} & \multicolumn{1}{l|}{\textbf{16/61}}                                             & \multicolumn{1}{l|}{}                  & \multicolumn{1}{l|}{\textbf{176/176}} & \multicolumn{1}{l|}{\textbf{152/162}} & \textbf{43/124}                                            \\ \cline{2-5} \cline{7-9} 
\multicolumn{1}{|l|}{}                                                                                        & \multicolumn{1}{l|}{\texttt{resolution/type errors}}               & \multicolumn{1}{l|}{14/73}          & \multicolumn{1}{l|}{\textbf{30/80}} & \multicolumn{1}{l|}{\textbf{15/61}}                                             & \multicolumn{1}{l|}{}                  & \multicolumn{1}{l|}{0/176}            & \multicolumn{1}{l|}{10/162}           & \textbf{66/124}                                            \\ \hline
\multicolumn{9}{|l|}{}                                                                                                                                                                                                                                                                                                                                                                                                                                                                                                        \\ \hline
\multicolumn{1}{|l|}{\multirow{6}{*}{\textbf{\begin{tabular}[c]{@{}l@{}}Verification\\ Errors\end{tabular}}}} & \multicolumn{1}{l|}{\texttt{loop invariant violation}}             & \multicolumn{1}{l|}{4/73}           & \multicolumn{1}{l|}{1/80}           & \multicolumn{1}{l|}{\textbf{23/61}}                                             & \multicolumn{1}{l|}{\multirow{6}{*}{}} & \multicolumn{1}{l|}{0/176}            & \multicolumn{1}{l|}{0/162}            & \textbf{10/124}                                            \\ \cline{2-5} \cline{7-9} 
\multicolumn{1}{|l|}{}                                                                                        & \multicolumn{1}{l|}{\texttt{postcondition might not hold}}         & \multicolumn{1}{l|}{3/73}           & \multicolumn{1}{l|}{1/80}           & \multicolumn{1}{l|}{\textbf{13/61}}                                             & \multicolumn{1}{l|}{}                  & \multicolumn{1}{l|}{0/176}            & \multicolumn{1}{l|}{0/162}            & 7/124                                                      \\ \cline{2-5} \cline{7-9} 
\multicolumn{1}{|l|}{}                                                                                        & \multicolumn{1}{l|}{\texttt{bound errors}}                         & \multicolumn{1}{l|}{0/73}           & \multicolumn{1}{l|}{1/80}           & \multicolumn{1}{l|}{2/61}                                                       & \multicolumn{1}{l|}{}                  & \multicolumn{1}{l|}{0/176}            & \multicolumn{1}{l|}{0/162}            & 5/124                                                      \\ \cline{2-5} \cline{7-9} 
\multicolumn{1}{|l|}{}                                                                                        & \multicolumn{1}{l|}{\texttt{index out of range}}                   & \multicolumn{1}{l|}{2/73}           & \multicolumn{1}{l|}{2/80}           & \multicolumn{1}{l|}{5/61}                                                       & \multicolumn{1}{l|}{}                  & \multicolumn{1}{l|}{0/176}            & \multicolumn{1}{l|}{0/162}            & 2/124                                                      \\ \cline{2-5} \cline{7-9} 
\multicolumn{1}{|l|}{}                                                                                        & \multicolumn{1}{l|}{\texttt{possible division by zero}}            & \multicolumn{1}{l|}{0/73}           & \multicolumn{1}{l|}{0/80}           & \multicolumn{1}{l|}{0/61}                                                       & \multicolumn{1}{l|}{}                  & \multicolumn{1}{l|}{0/176}            & \multicolumn{1}{l|}{0/162}            & 1/124                                                      \\ \cline{2-5} \cline{7-9} 
\multicolumn{1}{|l|}{}                                                                                        & \multicolumn{1}{l|}{\texttt{verification time out}}                & \multicolumn{1}{l|}{1/73}           & \multicolumn{1}{l|}{0/80}           & \multicolumn{1}{l|}{0/61}                                                       & \multicolumn{1}{l|}{}                  & \multicolumn{1}{l|}{0/176}            & \multicolumn{1}{l|}{0/162}            & 0/124                                                      \\ \hline
\end{tabular}
\end{table}
We manually analyzed the error logs for all unverified methods and identified two major types of errors. We then examined the error types and classified them into several subgroups. Table~\ref{tab:error_results_count} demonstrates the presence of different types of errors appeared in all unverified methods.

\subsubsection{\textbf{Compilation Errors:}} We found compilation errors when a Dafny method was not verified because of incorrect Dafny syntax or unknown types. These errors appeared in two forms (i) \texttt{parse errors}, having wrong Dafny syntax and (ii) \texttt{resolution/type errors}, when a method implementation contains usages of any unknown types. In our manual inspection, we observed that the majority of LLMs synthesis methods were not verified because of compilation errors. As shown in Table~\ref{tab:error_results_count}, in \textit{Contextless} and \textit{Signature} prompting, 100\% and 93.83\% of \palm synthesis methods were unverified because of \texttt{parse errors}. In \textit{Dynamic Few-Shot} RAG prompting, \texttt{parse errors} were minimized, nevertheless, \texttt{resolution/type errors} appeared in around 53.22\% of cases. We inspected the reasons for \texttt{resolution/type errors} and found that \palm generated Dafny methods assuming the presence of basic library methods/functions without providing any implementation. Consequently, the verifier raised a \texttt{resolution/type errors} because of the usages of unknown method/function.
\noindent
Similarly, \gpt also struggled generating correct Dafny syntax in \textit{Contextless} and \textit{Signature} prompting. In \textit{Contextless}, 51 out of 73 unverified methods have \texttt{parse errors}. It gradually overcome and improved its performance in \textit{Dynamic Few-Shot} prompting where only 16 out of 61 unverified methods had \texttt{parse errors}. Regarding, \texttt{resolution/type errors}, \gpt generated such errors most (30/80) in \textit{Signature} prompting whereas in \textit{Dynamic Few-Shot} prompting, \gpt had only 15 of such errors out of 61 unverified methods.
\vskip-0.5cm
\subsubsection{\textbf{Verification Errors:}} These types of errors occurred when a Dafny method has correct syntax but the specifications or implementation were incorrect. We found six different types of verification errors, shown in Table~\ref{tab:error_results_count}. Most of the syntactically correct unverified methods failed to verify because of \texttt{loop invariant violation} and/or \texttt{postcondition might not hold} errors. In \textit{Contextless} and \textit{Signature} prompting, there were very few verification errors since most of the methods had compilation errors. However, in \textit{Dynamic Few-Shot} prompting, \gpt had \texttt{loop invariant violation}  in 23 cases and \texttt{postcondition might not hold}  in 13 out of 61 unverified methods. We observed that these errors appeared for two reasons: (i) the specifications (postconditions or invariants) were correct but the implementation was incorrect to prove the specifications; and (ii)  the implementation was correct but the specifications were incorrect.
\vskip-0.2cm
\subsection{Summary of Findings}
\begin{tcolorbox}[width=\textwidth, colframe=black, colback=anti-flashwhite!30, boxsep=1mm, arc=1mm,bottom=0mm, left=3pt]

\textbf{RQ1~[Contextless Prompting]:} Our experiments
show that, given a simple problem description, LLMs can generate plausible Dafny methods,
although the resulting methods are not necessarily syntactically correct.
Specifically, GPT-4 is capable of generating Dafny methods, but often cannot generate the formal specifications. Without proper formal specifications, the internal verification annotations, when they exist, become moot.
\end{tcolorbox}
\vspace{-3mm}
\begin{tcolorbox}[width=\textwidth, colframe=black, colback=anti-flashwhite!30, boxsep=1mm, arc=1mm, bottom=0mm, left=3pt]

\textbf{RQ2~[Signature Prompting]:} 
Additional context --- method signature and test cases --- reduced the performance of both the LLMs that we evaluated. 
This contrasts with LLMs undertaking more classical code synthesis tasks, 
where additional contextual information generally improves their performance.
\end{tcolorbox}
\vspace{-3mm}
\begin{tcolorbox}[width=\textwidth, colframe=black, colback=anti-flashwhite!30, boxsep=1mm, arc=1mm, bottom=0mm, left=3pt]

\textbf{RQ3~[Dynamic Few-Shot Prompting]:}
Our results show that  \textit{Chain of Thought (CoT)} prompts, separately prompting for specifications, and then chaining those generated specifications in a second prompt with retrieval augmentation generated (i.e. RAG) semantically similar few-shot examples, significantly increased performance. Engineering these prompts led both LLMs to generate explicit pre- and post-conditions for all the test problems.
Specifically for \gpt, 100\% of all generated methods contained formal specifications, and 58\% of them (103/178) had strong (i.e. correct) specifications that successfully verified.
\end{tcolorbox}

\section{Related Work}
\label{sec:relwork}
\subsection{Program Synthesis and Verification with Dafny}
In the last 20 years, formal methods for software synthesis \cite{FTPL-synthesis} and verification \cite{FTPL-formalism} have
moved from an esoteric research topic~\cite{rustan1998extended} to a set of
increasingly practical tools, and from doctoral study
to undergraduate degrees~\cite{Jones2021}.  Tools such as Dafny, SAW, and SPIN are increasingly
mature enough to support industrial application~\cite{Greengard2021,wayne2018temporal}: 
indeed a  main barrier to adoption remains a lack of software
engineers trained in their use~\cite{Garavel2020}.
\par
Dafny is used in production to develop verified implementations  encryption algorithms 
\cite{DBLP:journals/iacr/YangWCCY23},  Ethereum Virtual Machine (EVM) bytecodes \cite{DBLP:conf/fm/CassezFGPQ23}, and even  quantum circuitry \cite{DBLP:journals/corr/abs-2211-06411}. Dafny is also the topic of ongoing research in verification and synthesis. 
\citet{DBLP:conf/issta/IrfanPRRT22} introduced XDsmith, a fuzzing differential testing framework that generates random Dafny programs with known verification output. XDsmith conducts differential testing on the Dafny compiler to evaluate Dafny's soundness and precision. 
In the Amazon Automated Reasoning group, \citet{DBLP:conf/tacas/ChakarovFRR22} developed an approach to generate counterexamples in Dafny syntax when the  Dafny verifier cannot establish a valid proof.
\par
This work testifies to the ongoing relevance of Dafny-style program verification in research, and increasingly in practice. Our results indicate that LLMs are likely to be a fruitful direction for assisting  with Dafny programming, in both research and practice.

\subsection{LLMs for Formal Methods}
\revise{
Researchers have been actively working on developing automated proof synthesis and theorem provers~\cite{DBLP:conf/pldi/Sanchez-SternAS20}. Several efficient proof synthesis tools have been developed based on pre-computed facts reasoning, heuristic search and identifiers, such as CoqHammer~\cite{DBLP:journals/jar/CzajkaK18}, TacTok~\cite{DBLP:journals/pacmpl/FirstBG20}, Passport \cite{DBLP:journals/toplas/SanchezSternFZKBR23}. Recently developed LLMs based approaches, for instance, ASTactic~\cite{DBLP:conf/icml/YangD19}, Diva~\cite{DBLP:conf/icse/FirstB22} Thor~\cite{DBLP:conf/nips/JiangLTCOMWJ22} and Baldur~\cite{DBLP:conf/sigsoft/FirstRRB23} outperform the previous approaches to synthesize whole or partial proofs and automatically prove theorems employing the interactive proof assistants Coq and Isabelle/HOL.}
\par
Apart from proof synthesis and theorem proving, LLMs have also been applied to translate natural language into formal specifications, proofs, and  theorems \cite{DBLP:conf/acl/QiaoO0CYDTHC23,DBLP:journals/corr/abs-2205-15231,DBLP:conf/acl/2023/LiMaking}. Wu et al.\ translated natural language problems into Isabelle/HOL \cite{DBLP:conf/nips/WuJLRSJS22}, while Cunningham et al.\ experimented generating proofs for code verification \cite{DBLP:journals/corr/abs-2301-02195}.  To enhance reasoning, Madaan et al.\ generated event graphs \cite{DBLP:conf/emnlp/MadaanZ0YN22} and found that LLMs trained on source code outperformed those trained only on text.
\par
Most contemporary LLMs have limitations when it comes to algorithmic reasoning
\cite{DBLP:journals/corr/abs-2108-07732} although this may be mitigated by careful prompting
\cite{DBLP:journals/corr/abs-2211-09066}.
While \gpt can generate plausible texts about mathematical problems,
it cannot currently deliver correct mathematical proofs or solutions \cite{DBLP:journals/corr/abs-2301-13867}. 
The Minerva model, however, is pre-trained on a mathematics corpus \cite{DBLP:conf/sigsoft/FirstRRB23}, and also aims to generate proofs in Isabelle/HOL. 
\par
Learning techniques are in vogue in software engineering research
\cite{DBLP:journals/tosem/WatsonCNMP22,DBLP:journals/csur/YangXLG22,DBLP:journals/csur/AllamanisBDS18}. LLMs have been evaluated across a range of tasks such as 
code generation, code completion, code summarization \cite{DBLP:conf/icse/TufanoPB23, nashid2023retrieval}, code repair \cite{DBLP:conf/icml/YasunagaL20,DBLP:journals/corr/abs-2205-00180,DBLP:conf/msr/MashhadiH21}, program transformation and synthesis \cite{DBLP:journals/corr/abs-2108-07732}, test generation \cite{DBLP:journals/tse/MastropaoloCNSPOB23,tufano2020unit}, bug fixing \cite{DBLP:conf/kbse/ChakrabortyR21,DBLP:journals/tosem/TufanoWBPWP19}, code review \cite{DBLP:conf/icse/TufanoMMPPB22,DBLP:conf/sigsoft/LiYJYLHLZ22} --- for some tasks, 
performing as well as or better than the state of the art 
\cite{DBLP:journals/tosem/WatsonCNMP22,DBLP:journals/csur/YangXLG22,DBLP:journals/csur/AllamanisBDS18}, especially when prompts are carefully constructed  \cite{nashid2023retrieval}.
Recent studies suggest that applying LLMs to code-related tasks has the potential 
to improve developer productivity \cite{nashid2023retrieval,DBLP:conf/kbse/ChakrabortyR21,DBLP:journals/tosem/TufanoWBPWP19}.
\par
All this work (both formal methods and programming tasks) focuses on careful training of LLMs in mathematics --- or otherwise extending their reasoning power. Our results explore an alternative approach: letting LLMs do what they are good at --- generating plausible solutions --- and then employing program verification techniques to do what they are good at --- verifying whether or not those solutions are correct.
\section{Threats to Validity}
\label{sec:threats}
Following \citet{siegmunds} and \citet{feldt2010validity} we  identified a number of threats to the validity of this study. The closed nature of our experiments and the straightforward decision criteria (can Dafny compile a program? verify a program?) should ensure construct and internal validity: we are interested in the mechanistic question of whether a contemporary language model can generate code for a verifiable programming language, and we test that directly. Our subjective judgments were engaged primarily to classify mistakes and assess the fitness of the postconditions given the natural language prompts. Our qualitative evaluation used standard inspection methods to reach consensus, and all our data, prompts, coding, and intermediate analyses are available. 
\par
External validity considers how well these results would generalize to other language models or  programming languages --- or even to different versions of the tools we evaluated.  In designing the study,  we were  aware that both the language model (GPT-4) and the programming language (Dafny) were evolving very rapidly: the version of GPT-4 we used had been available for only 14 days and Dafny 4.0 was released on 3rd March 2023; furthermore,  these versions embodied significant improvements over previous versions of these tools. The only sure assertions we can make about the \textit{detailed results} of this study are that they are unlikely to generalize: we expect the results will be different --- most likely,  better --- if we repeated the study with later versions of the language model or the programming language.  Unfortunately, 
there is no way to avoid these problems while studying contemporary software under continual, rapid, development:  we consider the advantages of using the latest contemporary versions outweighs any increase in reproducibility we could get by e.g.\ using earlier stable versions that are several years old. We have, however, mitigated this threat as well as possible. We have identified the precise versions of the software and hardware that we used. The research artifacts of this study are publicly available at the companion \href{https://github.com/Mondego/dafny-synthesis}{repository}.

\vspace{-2mm}
\section{Conclusion}
\label{sec:conclusion}
We investigated the potential of large language models (\gpt\ and \palm) to synthesize method specifications and bodies in the Dafny verification-aware programming language. We found that, given appropriate prompts, \gpt~can generate Dafny methods, the pre- and postconditions needed to specify those methods, and the internal annotations required for Dafny to verify the methods against the  specifications. Carefully engineered \textit{Chain of Thought (CoT)} prompts with retrieval augmentation generated semantically similar few-shot examples explaining problem decomposition step by step yielded the best performance, generating the most post-conditions and minimizing unnecessary preconditions. Of the 178 problems in our test dataset, \gpt  synthesized 103 methods with nontrivial specification that Dafny could verify:  many of the remaining methods exhibited only minor errors. These findings underscore the importance of careful prompt design, and suggest that with further improvements in prompt design and more examples, \gpt can serve as the core of a ``Programmer's Verification Apprentice'' for writing formally verified methods in Dafny. Although this work is  exploratory, the results demonstrate great potential for more focused research on program synthesis with LLMs. We hypothesize that incorporating the tools and techniques of program verification may be an equally fruitful direction for research into generative models more generally, whenever their output must not only be \textit{plausible} but also \textit{correct}.

\vspace{-2mm}
\section{Data Availability}
\label{sec:data-availability}
The complete artifacts of this study, including the evaluation benchmark, collection of verified Dafny code, 
prompts for each problem, LLMs’ response and our manual evaluation all are publicly available in the GitHub repository: \href{https://github.com/Mondego/dafny-synthesis/releases/tag/Artifacts%40FSE24-Reviewed}{\faGithub~dafny-synthesis} release
and this persistent  \href{https://doi.org/10.6084/m9.figshare.25999807}{DOI}.


\vspace{-2mm}
\bibliographystyle{ACM-Reference-Format}
\bibliography{reference}


\begin{thebibliography}{104}


\ifx \showCODEN    \undefined \def \showCODEN     #1{\unskip}     \fi
\ifx \showDOI      \undefined \def \showDOI       #1{#1}\fi
\ifx \showISBNx    \undefined \def \showISBNx     #1{\unskip}     \fi
\ifx \showISBNxiii \undefined \def \showISBNxiii  #1{\unskip}     \fi
\ifx \showISSN     \undefined \def \showISSN      #1{\unskip}     \fi
\ifx \showLCCN     \undefined \def \showLCCN      #1{\unskip}     \fi
\ifx \shownote     \undefined \def \shownote      #1{#1}          \fi
\ifx \showarticletitle \undefined \def \showarticletitle #1{#1}   \fi
\ifx \showURL      \undefined \def \showURL       {\relax}        \fi
\providecommand\bibfield[2]{#2}
\providecommand\bibinfo[2]{#2}
\providecommand\natexlab[1]{#1}
\providecommand\showeprint[2][]{arXiv:#2}

\bibitem[{Aakanksha Chowdhery et al.}(2022)]%
        {DBLP:journals/corr/abs-2204-02311}
\bibfield{author}{\bibinfo{person}{{Aakanksha Chowdhery et al.}}} \bibinfo{year}{2022}\natexlab{}.
\newblock \showarticletitle{PaLM: Scaling Language Modeling with Pathways}.
\newblock \bibinfo{journal}{\emph{CoRR}}  \bibinfo{volume}{abs/2204.02311} (\bibinfo{year}{2022}).
\newblock


\bibitem[Allamanis et~al\mbox{.}(2018)]%
        {DBLP:journals/csur/AllamanisBDS18}
\bibfield{author}{\bibinfo{person}{Miltiadis Allamanis}, \bibinfo{person}{Earl~T. Barr}, \bibinfo{person}{Premkumar~T. Devanbu}, {and} \bibinfo{person}{Charles Sutton}.} \bibinfo{year}{2018}\natexlab{}.
\newblock \showarticletitle{A Survey of Machine Learning for Big Code and Naturalness}.
\newblock \bibinfo{journal}{\emph{{ACM} Comput. Surv.}} \bibinfo{volume}{51}, \bibinfo{number}{4} (\bibinfo{year}{2018}), \bibinfo{pages}{81:1--81:37}.
\newblock
\urldef\tempurl%
\url{https://doi.org/10.1145/3212695}
\showURL{%
\tempurl}


\bibitem[Amazon(2023)]%
        {awsautomatedreasoning}
\bibfield{author}{\bibinfo{person}{Amazon}.} \bibinfo{year}{2023}\natexlab{}.
\newblock \bibinfo{title}{Automated reasoning}.
\newblock \bibinfo{howpublished}{\url{https://www.amazon.science/research-areas/automated-reasoning}}.
\newblock
\newblock
\shownote{[Online], [Accessed: 2023-09-20]}.


\bibitem[Austin et~al\mbox{.}(2021)]%
        {DBLP:journals/corr/abs-2108-07732}
\bibfield{author}{\bibinfo{person}{Jacob Austin}, \bibinfo{person}{Augustus Odena}, \bibinfo{person}{Maxwell~I. Nye}, \bibinfo{person}{Maarten Bosma}, \bibinfo{person}{Henryk Michalewski}, \bibinfo{person}{David Dohan}, \bibinfo{person}{Ellen Jiang}, \bibinfo{person}{Carrie~J. Cai}, \bibinfo{person}{Michael Terry}, \bibinfo{person}{Quoc~V. Le}, {and} \bibinfo{person}{Charles Sutton}.} \bibinfo{year}{2021}\natexlab{}.
\newblock \showarticletitle{Program Synthesis with Large Language Models}.
\newblock \bibinfo{journal}{\emph{CoRR}}  \bibinfo{volume}{abs/2108.07732} (\bibinfo{year}{2021}).
\newblock
\urldef\tempurl%
\url{https://arxiv.org/abs/2108.07732}
\showURL{%
\tempurl}


\bibitem[Azaria and Mitchell(2023)]%
        {azaria2023internal}
\bibfield{author}{\bibinfo{person}{Amos Azaria} {and} \bibinfo{person}{Tom~M. Mitchell}.} \bibinfo{year}{2023}\natexlab{}.
\newblock \showarticletitle{The Internal State of an {LLM} Knows When It's Lying}. In \bibinfo{booktitle}{\emph{Findings of the Association for Computational Linguistics: {EMNLP} 2023, Singapore, December 6-10, 2023}}, \bibfield{editor}{\bibinfo{person}{Houda Bouamor}, \bibinfo{person}{Juan Pino}, {and} \bibinfo{person}{Kalika Bali}} (Eds.). \bibinfo{publisher}{Association for Computational Linguistics}, \bibinfo{pages}{967--976}.
\newblock
\urldef\tempurl%
\url{https://aclanthology.org/2023.findings-emnlp.68}
\showURL{%
\tempurl}


\bibitem[Bach et~al\mbox{.}(2022)]%
        {bach2022}
\bibfield{author}{\bibinfo{person}{Stephen~H. Bach}, \bibinfo{person}{Victor Sanh}, \bibinfo{person}{Zheng~Xin Yong}, \bibinfo{person}{Albert Webson}, \bibinfo{person}{Colin Raffel}, \bibinfo{person}{Nihal~V. Nayak}, \bibinfo{person}{Abheesht Sharma}, \bibinfo{person}{Taewoon Kim}, \bibinfo{person}{M.~Saiful Bari}, \bibinfo{person}{Thibault F{\'{e}}vry}, \bibinfo{person}{Zaid Alyafeai}, \bibinfo{person}{Manan Dey}, \bibinfo{person}{Andrea Santilli}, \bibinfo{person}{Zhiqing Sun}, \bibinfo{person}{Srulik Ben{-}David}, \bibinfo{person}{Canwen Xu}, \bibinfo{person}{Gunjan Chhablani}, \bibinfo{person}{Han Wang}, \bibinfo{person}{Jason~Alan Fries}, \bibinfo{person}{Maged~Saeed AlShaibani}, \bibinfo{person}{Shanya Sharma}, \bibinfo{person}{Urmish Thakker}, \bibinfo{person}{Khalid Almubarak}, \bibinfo{person}{Xiangru Tang}, \bibinfo{person}{Dragomir~R. Radev}, \bibinfo{person}{Mike~Tian{-}Jian Jiang}, {and} \bibinfo{person}{Alexander~M. Rush}.} \bibinfo{year}{2022}\natexlab{}.
\newblock \showarticletitle{PromptSource: An Integrated Development Environment and Repository for Natural Language Prompts}. In \bibinfo{booktitle}{\emph{Proceedings of the 60th Annual Meeting of the Association for Computational Linguistics, {ACL} 2022 - System Demonstrations, Dublin, Ireland, May 22-27, 2022}}, \bibfield{editor}{\bibinfo{person}{Valerio Basile}, \bibinfo{person}{Zornitsa Kozareva}, {and} \bibinfo{person}{Sanja Stajner}} (Eds.). \bibinfo{publisher}{Association for Computational Linguistics}, \bibinfo{pages}{93--104}.
\newblock
\urldef\tempurl%
\url{https://doi.org/10.18653/V1/2022.ACL-DEMO.9}
\showDOI{\tempurl}


\bibitem[Barrett et~al\mbox{.}(2021)]%
        {SMT-Handbook2021}
\bibfield{author}{\bibinfo{person}{Clark~W. Barrett}, \bibinfo{person}{Roberto Sebastiani}, \bibinfo{person}{Sanjit~A. Seshia}, {and} \bibinfo{person}{Cesare Tinelli}.} \bibinfo{year}{2021}\natexlab{}.
\newblock \showarticletitle{Satisfiability Modulo Theories}.
\newblock In \bibinfo{booktitle}{\emph{Handbook of Satisfiability - Second Edition}}, \bibfield{editor}{\bibinfo{person}{Armin Biere}, \bibinfo{person}{Marijn Heule}, \bibinfo{person}{Hans van Maaren}, {and} \bibinfo{person}{Toby Walsh}} (Eds.). \bibinfo{series}{Frontiers in Artificial Intelligence and Applications}, Vol.~\bibinfo{volume}{336}. \bibinfo{publisher}{{IOS} Press}, \bibinfo{pages}{1267--1329}.
\newblock
\urldef\tempurl%
\url{https://doi.org/10.3233/FAIA201017}
\showDOI{\tempurl}


\bibitem[Beurer{-}Kellner et~al\mbox{.}(2023)]%
        {promptgramming2023}
\bibfield{author}{\bibinfo{person}{Luca Beurer{-}Kellner}, \bibinfo{person}{Marc Fischer}, {and} \bibinfo{person}{Martin~T. Vechev}.} \bibinfo{year}{2023}\natexlab{}.
\newblock \showarticletitle{Prompting Is Programming: {A} Query Language for Large Language Models}.
\newblock \bibinfo{journal}{\emph{Proc. {ACM} Program. Lang.}} \bibinfo{volume}{7}, \bibinfo{number}{{PLDI}} (\bibinfo{year}{2023}), \bibinfo{pages}{1946--1969}.
\newblock
\urldef\tempurl%
\url{https://doi.org/10.1145/3591300}
\showDOI{\tempurl}


\bibitem[Cassez et~al\mbox{.}(2023)]%
        {DBLP:conf/fm/CassezFGPQ23}
\bibfield{author}{\bibinfo{person}{Franck Cassez}, \bibinfo{person}{Joanne Fuller}, \bibinfo{person}{Milad~K. Ghale}, \bibinfo{person}{David~J. Pearce}, {and} \bibinfo{person}{Horacio Mijail~Anton Quiles}.} \bibinfo{year}{2023}\natexlab{}.
\newblock \showarticletitle{Formal and Executable Semantics of the Ethereum Virtual Machine in Dafny}. In \bibinfo{booktitle}{\emph{Formal Methods - 25th International Symposium, {FM} 2023, L{\"{u}}beck, Germany, March 6-10, 2023, Proceedings}} \emph{(\bibinfo{series}{Lecture Notes in Computer Science}, Vol.~\bibinfo{volume}{14000})}, \bibfield{editor}{\bibinfo{person}{Marsha Chechik}, \bibinfo{person}{Joost{-}Pieter Katoen}, {and} \bibinfo{person}{Martin Leucker}} (Eds.). \bibinfo{publisher}{Springer}, \bibinfo{pages}{571--583}.
\newblock
\urldef\tempurl%
\url{https://doi.org/10.1007/978-3-031-27481-7\_32}
\showDOI{\tempurl}


\bibitem[Chakarov et~al\mbox{.}(2022)]%
        {DBLP:conf/tacas/ChakarovFRR22}
\bibfield{author}{\bibinfo{person}{Aleksandar Chakarov}, \bibinfo{person}{Aleksandr Fedchin}, \bibinfo{person}{Zvonimir Rakamaric}, {and} \bibinfo{person}{Neha Rungta}.} \bibinfo{year}{2022}\natexlab{}.
\newblock \showarticletitle{Better Counterexamples for Dafny}. In \bibinfo{booktitle}{\emph{{TACAS}}} \emph{(\bibinfo{series}{Lecture Notes in Computer Science}, Vol.~\bibinfo{volume}{13243})}, \bibfield{editor}{\bibinfo{person}{Dana Fisman} {and} \bibinfo{person}{Grigore Rosu}} (Eds.). \bibinfo{publisher}{Springer}, \bibinfo{pages}{404--411}.
\newblock
\urldef\tempurl%
\url{https://doi.org/10.1007/978-3-030-99524-9\_23}
\showURL{%
\tempurl}


\bibitem[Chakraborty and Ray(2021)]%
        {DBLP:conf/kbse/ChakrabortyR21}
\bibfield{author}{\bibinfo{person}{Saikat Chakraborty} {and} \bibinfo{person}{Baishakhi Ray}.} \bibinfo{year}{2021}\natexlab{}.
\newblock \showarticletitle{On Multi-Modal Learning of Editing Source Code}. In \bibinfo{booktitle}{\emph{36th {IEEE/ACM} International Conference on Automated Software Engineering, {ASE} 2021, Melbourne, Australia, November 15-19, 2021}}. \bibinfo{publisher}{{IEEE}}, \bibinfo{pages}{443--455}.
\newblock
\urldef\tempurl%
\url{https://doi.org/10.1109/ASE51524.2021.9678559}
\showURL{%
\tempurl}


\bibitem[Chen et~al\mbox{.}(2023)]%
        {DBLP:journals/corr/abs-2302-12692}
\bibfield{author}{\bibinfo{person}{Zekai Chen}, \bibinfo{person}{Mariann Micsinai~Balan}, {and} \bibinfo{person}{Kevin Brown}.} \bibinfo{year}{2023}\natexlab{}.
\newblock \showarticletitle{Boosting Transformers and Language Models for Clinical Prediction in Immunotherapy}. In \bibinfo{booktitle}{\emph{Proceedings of the 61st Annual Meeting of the Association for Computational Linguistics (Volume 5: Industry Track)}}, \bibfield{editor}{\bibinfo{person}{Sunayana Sitaram}, \bibinfo{person}{Beata Beigman~Klebanov}, {and} \bibinfo{person}{Jason~D Williams}} (Eds.). \bibinfo{publisher}{Association for Computational Linguistics}, \bibinfo{address}{Toronto, Canada}, \bibinfo{pages}{332--340}.
\newblock
\urldef\tempurl%
\url{https://doi.org/10.18653/v1/2023.acl-industry.32}
\showDOI{\tempurl}


\bibitem[CodeWhisperer(2023)]%
        {codewhisperer}
\bibfield{author}{\bibinfo{person}{Amazon CodeWhisperer}.} \bibinfo{year}{2023}\natexlab{}.
\newblock \bibinfo{title}{Amazon CodeWhisperer}.
\newblock \bibinfo{howpublished}{\url{https://aws.amazon.com/codewhisperer/}}.
\newblock
\newblock
\shownote{[Online], [Accessed: 2023-09-20]}.


\bibitem[CompCert(2023)]%
        {CompCert}
\bibfield{author}{\bibinfo{person}{CompCert}.} \bibinfo{year}{2023}\natexlab{}.
\newblock \bibinfo{title}{CompCert}.
\newblock \bibinfo{howpublished}{\url{https://github.com/AbsInt/CompCert}}.
\newblock
\newblock
\shownote{[Online], [Accessed: 2023-09-20]}.


\bibitem[Cook(2018)]%
        {DBLP:conf/cav/Cook18}
\bibfield{author}{\bibinfo{person}{Byron Cook}.} \bibinfo{year}{2018}\natexlab{}.
\newblock \showarticletitle{Formal Reasoning About the Security of Amazon Web Services}. In \bibinfo{booktitle}{\emph{Computer Aided Verification - 30th International Conference, {CAV} 2018, Held as Part of the Federated Logic Conference, FloC 2018, Oxford, UK, July 14-17, 2018, Proceedings, Part {I}}} \emph{(\bibinfo{series}{Lecture Notes in Computer Science}, Vol.~\bibinfo{volume}{10981})}, \bibfield{editor}{\bibinfo{person}{Hana Chockler} {and} \bibinfo{person}{Georg Weissenbacher}} (Eds.). \bibinfo{publisher}{Springer}, \bibinfo{pages}{38--47}.
\newblock
\urldef\tempurl%
\url{https://doi.org/10.1007/978-3-319-96145-3\_3}
\showDOI{\tempurl}


\bibitem[Copilot(2023)]%
        {copilot}
\bibfield{author}{\bibinfo{person}{GitHub Copilot}.} \bibinfo{year}{2023}\natexlab{}.
\newblock \bibinfo{title}{GitHub Copilot}.
\newblock \bibinfo{howpublished}{\url{https://github.com/features/copilot}}.
\newblock
\newblock
\shownote{[Online], [Accessed: 2023-09-20]}.


\bibitem[Cunningham et~al\mbox{.}(2022)]%
        {DBLP:journals/corr/abs-2301-02195}
\bibfield{author}{\bibinfo{person}{Garett Cunningham}, \bibinfo{person}{Razvan~C. Bunescu}, {and} \bibinfo{person}{David Juedes}.} \bibinfo{year}{2022}\natexlab{}.
\newblock \showarticletitle{Towards Autoformalization of Mathematics and Code Correctness: Experiments with Elementary Proofs}.
\newblock  (\bibinfo{year}{2022}), \bibinfo{pages}{25--32}.
\newblock
\urldef\tempurl%
\url{https://aclanthology.org/2022.mathnlp-1.4.pdf}
\showURL{%
\tempurl}


\bibitem[Czajka and Kaliszyk(2018)]%
        {DBLP:journals/jar/CzajkaK18}
\bibfield{author}{\bibinfo{person}{Lukasz Czajka} {and} \bibinfo{person}{Cezary Kaliszyk}.} \bibinfo{year}{2018}\natexlab{}.
\newblock \showarticletitle{Hammer for Coq: Automation for Dependent Type Theory}.
\newblock \bibinfo{journal}{\emph{J. Autom. Reason.}} \bibinfo{volume}{61}, \bibinfo{number}{1-4} (\bibinfo{year}{2018}), \bibinfo{pages}{423--453}.
\newblock
\urldef\tempurl%
\url{https://doi.org/10.1007/S10817-018-9458-4}
\showDOI{\tempurl}


\bibitem[Dafny(2023)]%
        {dafny-github}
\bibfield{author}{\bibinfo{person}{Dafny}.} \bibinfo{year}{2023}\natexlab{}.
\newblock \bibinfo{title}{dafny-lang}.
\newblock \bibinfo{howpublished}{\url{https://github.com/dafny-lang/dafny}}.
\newblock
\newblock
\shownote{[Online], [Accessed: 2023-09-20]}.


\bibitem[Dang et~al\mbox{.}(2022)]%
        {dang2022prompt}
\bibfield{author}{\bibinfo{person}{Hai Dang}, \bibinfo{person}{Lukas Mecke}, \bibinfo{person}{Florian Lehmann}, \bibinfo{person}{Sven Goller}, {and} \bibinfo{person}{Daniel Buschek}.} \bibinfo{year}{2022}\natexlab{}.
\newblock \showarticletitle{How to Prompt? Opportunities and Challenges of Zero- and Few-Shot Learning for Human-AI Interaction in Creative Applications of Generative Models}.
\newblock \bibinfo{journal}{\emph{CoRR}}  \bibinfo{volume}{abs/2209.01390}.
\newblock
\urldef\tempurl%
\url{https://doi.org/10.48550/arxiv.2209.01390}
\showURL{%
\tempurl}


\bibitem[de~Moura and Bj{\o}rner(2008)]%
        {DBLP:conf/tacas/MouraB08}
\bibfield{author}{\bibinfo{person}{Leonardo~Mendon{\c{c}}a de Moura} {and} \bibinfo{person}{Nikolaj~S. Bj{\o}rner}.} \bibinfo{year}{2008}\natexlab{}.
\newblock \showarticletitle{{Z3:} An Efficient {SMT} Solver}. In \bibinfo{booktitle}{\emph{{TACAS}}}. \bibinfo{pages}{337--340}.
\newblock
\urldef\tempurl%
\url{https://doi.org/10.1007/978-3-540-78800-3\_24}
\showURL{%
\tempurl}


\bibitem[Deckers et~al\mbox{.}(2023)]%
        {deckers2023}
\bibfield{author}{\bibinfo{person}{Niklas Deckers}, \bibinfo{person}{Maik Fr{\"{o}}be}, \bibinfo{person}{Johannes Kiesel}, \bibinfo{person}{Gianluca Pandolfo}, \bibinfo{person}{Christopher Schr{\"{o}}der}, \bibinfo{person}{Benno Stein}, {and} \bibinfo{person}{Martin Potthast}.} \bibinfo{year}{2023}\natexlab{}.
\newblock \showarticletitle{The Infinite Index: Information Retrieval on Generative Text-To-Image Models}. In \bibinfo{booktitle}{\emph{Proceedings of the 2023 Conference on Human Information Interaction and Retrieval, {CHIIR} 2023, Austin, TX, USA, March 19-23, 2023}}, \bibfield{editor}{\bibinfo{person}{Jacek Gwizdka} {and} \bibinfo{person}{Soo~Young Rieh}} (Eds.). \bibinfo{publisher}{{ACM}}, \bibinfo{pages}{172--186}.
\newblock
\urldef\tempurl%
\url{https://doi.org/10.1145/3576840.3578327}
\showDOI{\tempurl}


\bibitem[Dohan et~al\mbox{.}(2022)]%
        {dohan2022language}
\bibfield{author}{\bibinfo{person}{David Dohan}, \bibinfo{person}{Winnie Xu}, \bibinfo{person}{Aitor Lewkowycz}, \bibinfo{person}{Jacob Austin}, \bibinfo{person}{David Bieber}, \bibinfo{person}{Raphael~Gontijo Lopes}, \bibinfo{person}{Yuhuai Wu}, \bibinfo{person}{Henryk Michalewski}, \bibinfo{person}{Rif~A. Saurous}, \bibinfo{person}{Jascha Sohl{-}Dickstein}, \bibinfo{person}{Kevin Murphy}, {and} \bibinfo{person}{Charles Sutton}.} \bibinfo{year}{2022}\natexlab{}.
\newblock \bibinfo{title}{Language Model Cascades}.
\newblock
\newblock
\urldef\tempurl%
\url{https://doi.org/10.48550/arXiv.2207.10342}
\showURL{%
\tempurl}


\bibitem[Erbsen et~al\mbox{.}(2020)]%
        {DBLP:journals/sigops/ErbsenPGSC20}
\bibfield{author}{\bibinfo{person}{Andres Erbsen}, \bibinfo{person}{Jade Philipoom}, \bibinfo{person}{Jason Gross}, \bibinfo{person}{Robert Sloan}, {and} \bibinfo{person}{Adam Chlipala}.} \bibinfo{year}{2020}\natexlab{}.
\newblock \showarticletitle{Simple High-Level Code For Cryptographic Arithmetic: With Proofs, Without Compromises}.
\newblock \bibinfo{journal}{\emph{{ACM} {SIGOPS} Oper. Syst. Rev.}} \bibinfo{volume}{54}, \bibinfo{number}{1} (\bibinfo{year}{2020}), \bibinfo{pages}{23--30}.
\newblock
\urldef\tempurl%
\url{https://doi.org/10.1145/3421473.3421477}
\showURL{%
\tempurl}


\bibitem[Faria and Abreu(2023)]%
        {DBLP:conf/fsen/FariaA23}
\bibfield{author}{\bibinfo{person}{Jo{\~{a}}o~Pascoal Faria} {and} \bibinfo{person}{Rui Abreu}.} \bibinfo{year}{2023}\natexlab{}.
\newblock \showarticletitle{Case Studies of Development of Verified Programs with Dafny for Accessibility Assessment}. In \bibinfo{booktitle}{\emph{Fundamentals of Software Engineering - 10th International Conference, {FSEN} 2023, Tehran, Iran, May 4-5, 2023, Revised Selected Papers}} \emph{(\bibinfo{series}{Lecture Notes in Computer Science}, Vol.~\bibinfo{volume}{14155})}, \bibfield{editor}{\bibinfo{person}{Hossein Hojjat} {and} \bibinfo{person}{Erika {\'{A}}brah{\'{a}}m}} (Eds.). \bibinfo{publisher}{Springer}, \bibinfo{pages}{25--39}.
\newblock
\urldef\tempurl%
\url{https://doi.org/10.1007/978-3-031-42441-0\_3}
\showDOI{\tempurl}


\bibitem[Feldt and Magazinius(2010)]%
        {feldt2010validity}
\bibfield{author}{\bibinfo{person}{Robert Feldt} {and} \bibinfo{person}{Ana Magazinius}.} \bibinfo{year}{2010}\natexlab{}.
\newblock \showarticletitle{Validity Threats in Empirical Software Engineering Research - An Initial Survey}. In \bibinfo{booktitle}{\emph{Proceedings of the 22nd International Conference on Software Engineering {\&} Knowledge Engineering (SEKE'2010), Redwood City, San Francisco Bay, CA, USA, July 1 - July 3, 2010}}. \bibinfo{publisher}{Knowledge Systems Institute Graduate School}, \bibinfo{pages}{374--379}.
\newblock


\bibitem[First and Brun(2022)]%
        {DBLP:conf/icse/FirstB22}
\bibfield{author}{\bibinfo{person}{Emily First} {and} \bibinfo{person}{Yuriy Brun}.} \bibinfo{year}{2022}\natexlab{}.
\newblock \showarticletitle{Diversity-Driven Automated Formal Verification}. In \bibinfo{booktitle}{\emph{44th {IEEE/ACM} 44th International Conference on Software Engineering, {ICSE} 2022, Pittsburgh, PA, USA, May 25-27, 2022}}. \bibinfo{publisher}{{ACM}}, \bibinfo{pages}{1--13}.
\newblock
\urldef\tempurl%
\url{https://doi.org/10.1145/3510003.3510138}
\showDOI{\tempurl}


\bibitem[First et~al\mbox{.}(2020)]%
        {DBLP:journals/pacmpl/FirstBG20}
\bibfield{author}{\bibinfo{person}{Emily First}, \bibinfo{person}{Yuriy Brun}, {and} \bibinfo{person}{Arjun Guha}.} \bibinfo{year}{2020}\natexlab{}.
\newblock \showarticletitle{TacTok: semantics-aware proof synthesis}.
\newblock \bibinfo{journal}{\emph{Proc. {ACM} Program. Lang.}} \bibinfo{volume}{4}, \bibinfo{number}{{OOPSLA}} (\bibinfo{year}{2020}), \bibinfo{pages}{231:1--231:31}.
\newblock
\urldef\tempurl%
\url{https://doi.org/10.1145/3428299}
\showDOI{\tempurl}


\bibitem[First et~al\mbox{.}(2023)]%
        {DBLP:conf/sigsoft/FirstRRB23}
\bibfield{author}{\bibinfo{person}{Emily First}, \bibinfo{person}{Markus~N. Rabe}, \bibinfo{person}{Talia Ringer}, {and} \bibinfo{person}{Yuriy Brun}.} \bibinfo{year}{2023}\natexlab{}.
\newblock \showarticletitle{Baldur: Whole-Proof Generation and Repair with Large Language Models}. In \bibinfo{booktitle}{\emph{Proceedings of the 31st {ACM} Joint European Software Engineering Conference and Symposium on the Foundations of Software Engineering, {ESEC/FSE} 2023, San Francisco, CA, USA, December 3-9, 2023}}, \bibfield{editor}{\bibinfo{person}{Satish Chandra}, \bibinfo{person}{Kelly Blincoe}, {and} \bibinfo{person}{Paolo Tonella}} (Eds.). \bibinfo{publisher}{{ACM}}, \bibinfo{pages}{1229--1241}.
\newblock
\urldef\tempurl%
\url{https://doi.org/10.1145/3611643.3616243}
\showDOI{\tempurl}


\bibitem[Fran{\c{c}}a et~al\mbox{.}(2011)]%
        {airbusC2011}
\bibfield{author}{\bibinfo{person}{Ricardo~Bedin Fran{\c{c}}a}, \bibinfo{person}{Denis Favre{-}Felix}, \bibinfo{person}{Xavier Leroy}, \bibinfo{person}{Marc Pantel}, {and} \bibinfo{person}{Jean Souyris}.} \bibinfo{year}{2011}\natexlab{}.
\newblock \showarticletitle{Towards Formally Verified Optimizing Compilation in Flight Control Software}. In \bibinfo{booktitle}{\emph{Bringing Theory to Practice: Predictability and Performance in Embedded Systems, {DATE} Workshop {PPES} 2011, March 18, 2011, Grenoble, France}} \emph{(\bibinfo{series}{OASIcs}, Vol.~\bibinfo{volume}{18})}, \bibfield{editor}{\bibinfo{person}{Philipp Lucas}, \bibinfo{person}{Lothar Thiele}, \bibinfo{person}{Benoit Triquet}, \bibinfo{person}{Theo Ungerer}, {and} \bibinfo{person}{Reinhard Wilhelm}} (Eds.). \bibinfo{publisher}{Schloss Dagstuhl - Leibniz-Zentrum fuer Informatik, Germany}, \bibinfo{pages}{59--68}.
\newblock
\urldef\tempurl%
\url{https://doi.org/10.4230/OASICS.PPES.2011.59}
\showDOI{\tempurl}


\bibitem[Frieder et~al\mbox{.}(2023)]%
        {DBLP:journals/corr/abs-2301-13867}
\bibfield{author}{\bibinfo{person}{Simon Frieder}, \bibinfo{person}{Luca Pinchetti}, \bibinfo{person}{Ryan{-}Rhys Griffiths}, \bibinfo{person}{Tommaso Salvatori}, \bibinfo{person}{Thomas Lukasiewicz}, \bibinfo{person}{Philipp~Christian Petersen}, \bibinfo{person}{Alexis Chevalier}, {and} \bibinfo{person}{Julius Berner}.} \bibinfo{year}{2023}\natexlab{}.
\newblock \showarticletitle{Mathematical Capabilities of ChatGPT}.
\newblock \bibinfo{journal}{\emph{CoRR}}  \bibinfo{volume}{abs/2301.13867} (\bibinfo{year}{2023}).
\newblock
\urldef\tempurl%
\url{https://doi.org/10.48550/arXiv.2301.13867}
\showURL{%
\tempurl}


\bibitem[Galitsky(2023)]%
        {truth-O-meter-PP23}
\bibfield{author}{\bibinfo{person}{Boris~A. Galitsky}.} \bibinfo{year}{2023}\natexlab{}.
\newblock \showarticletitle{{Truth-O-Meter:} Collaborating with {LLM} in Fighting its Hallucinations}.
\newblock \bibinfo{journal}{\emph{Preprints}} (\bibinfo{year}{2023}).
\newblock
\urldef\tempurl%
\url{https://doi.org/10.20944/preprints202307.1723.v1}
\showDOI{\tempurl}


\bibitem[Garavel et~al\mbox{.}(2020)]%
        {Garavel2020}
\bibfield{author}{\bibinfo{person}{Hubert Garavel}, \bibinfo{person}{Maurice~H Ter~Beek}, {and} \bibinfo{person}{Jaco Van De~Pol}.} \bibinfo{year}{2020}\natexlab{}.
\newblock \showarticletitle{The 2020 expert survey on formal methods}. In \bibinfo{booktitle}{\emph{International Conference on Formal Methods for Industrial Critical Systems}}. Springer, \bibinfo{pages}{3--69}.
\newblock


\bibitem[GCC(2023)]%
        {GCC}
\bibfield{author}{\bibinfo{person}{GCC}.} \bibinfo{year}{2023}\natexlab{}.
\newblock \bibinfo{title}{GCC, the GNU Compiler Collection}.
\newblock \bibinfo{howpublished}{\url{https://gcc.gnu.org/}}.
\newblock
\newblock
\shownote{[Online], [Accessed: 2023-09-20]}.


\bibitem[Goues et~al\mbox{.}(2011)]%
        {le2011boogie}
\bibfield{author}{\bibinfo{person}{Claire~Le Goues}, \bibinfo{person}{K.~Rustan~M. Leino}, {and} \bibinfo{person}{Michal Moskal}.} \bibinfo{year}{2011}\natexlab{}.
\newblock \showarticletitle{The Boogie Verification Debugger (Tool Paper)}.
\newblock   \bibinfo{volume}{7041} (\bibinfo{year}{2011}), \bibinfo{pages}{407--414}.
\newblock
\urldef\tempurl%
\url{https://doi.org/10.1007/978-3-642-24690-6\_28}
\showDOI{\tempurl}


\bibitem[Greengard(2021)]%
        {Greengard2021}
\bibfield{author}{\bibinfo{person}{Samuel Greengard}.} \bibinfo{year}{2021}\natexlab{}.
\newblock \bibinfo{booktitle}{\emph{The Internet of Things}}.
\newblock \bibinfo{publisher}{MIT press}.
\newblock


\bibitem[Gulwani et~al\mbox{.}(2017)]%
        {FTPL-synthesis}
\bibfield{author}{\bibinfo{person}{Sumit Gulwani}, \bibinfo{person}{Oleksandr Polozov}, {and} \bibinfo{person}{Rishabh Singh}.} \bibinfo{year}{2017}\natexlab{}.
\newblock \showarticletitle{Program Synthesis}.
\newblock \bibinfo{journal}{\emph{{Founds.\ Trends.\ Prog.\ Lang.}}} \bibinfo{volume}{4}, \bibinfo{number}{1-2} (\bibinfo{year}{2017}), \bibinfo{pages}{1--119}.
\newblock
\urldef\tempurl%
\url{https://doi.org/10.1561/2500000010}
\showDOI{\tempurl}


\bibitem[Hallgren(2012)]%
        {hallgren2012computing}
\bibfield{author}{\bibinfo{person}{Kevin~A Hallgren}.} \bibinfo{year}{2012}\natexlab{}.
\newblock \showarticletitle{Computing inter-rater reliability for observational data: an overview and tutorial}.
\newblock \bibinfo{journal}{\emph{Tutorials in quantitative methods for psychology}} \bibinfo{volume}{8}, \bibinfo{number}{1} (\bibinfo{year}{2012}), \bibinfo{pages}{23}.
\newblock
\urldef\tempurl%
\url{https://doi.org/10.20982/tqmp.08.1.p023}
\showDOI{\tempurl}


\bibitem[Hawblitzel et~al\mbox{.}(2015)]%
        {DBLP:conf/sosp/HawblitzelHKLPR15}
\bibfield{author}{\bibinfo{person}{Chris Hawblitzel}, \bibinfo{person}{Jon Howell}, \bibinfo{person}{Manos Kapritsos}, \bibinfo{person}{Jacob~R. Lorch}, \bibinfo{person}{Bryan Parno}, \bibinfo{person}{Michael~L. Roberts}, \bibinfo{person}{Srinath T.~V. Setty}, {and} \bibinfo{person}{Brian Zill}.} \bibinfo{year}{2015}\natexlab{}.
\newblock \showarticletitle{IronFleet: proving practical distributed systems correct}. In \bibinfo{booktitle}{\emph{{SOSP}}}, \bibfield{editor}{\bibinfo{person}{Ethan~L. Miller} {and} \bibinfo{person}{Steven Hand}} (Eds.). \bibinfo{publisher}{{ACM}}, \bibinfo{pages}{1--17}.
\newblock
\urldef\tempurl%
\url{https://doi.org/10.1145/2815400.2815428}
\showURL{%
\tempurl}


\bibitem[Hawblitzel et~al\mbox{.}(2014)]%
        {DBLP:conf/osdi/HawblitzelHLNPZZ14}
\bibfield{author}{\bibinfo{person}{Chris Hawblitzel}, \bibinfo{person}{Jon Howell}, \bibinfo{person}{Jacob~R. Lorch}, \bibinfo{person}{Arjun Narayan}, \bibinfo{person}{Bryan Parno}, \bibinfo{person}{Danfeng Zhang}, {and} \bibinfo{person}{Brian Zill}.} \bibinfo{year}{2014}\natexlab{}.
\newblock \showarticletitle{Ironclad Apps: End-to-End Security via Automated Full-System Verification}. In \bibinfo{booktitle}{\emph{11th {USENIX} Symposium on Operating Systems Design and Implementation, {OSDI} '14, Broomfield, CO, USA, October 6-8, 2014}}, \bibfield{editor}{\bibinfo{person}{Jason Flinn} {and} \bibinfo{person}{Hank Levy}} (Eds.). \bibinfo{publisher}{{USENIX} Association}, \bibinfo{pages}{165--181}.
\newblock
\urldef\tempurl%
\url{https://www.usenix.org/conference/osdi14/technical-sessions/presentation/hawblitzel}
\showURL{%
\tempurl}


\bibitem[Hoare(1969)]%
        {DBLP:journals/cacm/Hoare69}
\bibfield{author}{\bibinfo{person}{C.~A.~R. Hoare}.} \bibinfo{year}{1969}\natexlab{}.
\newblock \showarticletitle{An Axiomatic Basis for Computer Programming}.
\newblock \bibinfo{journal}{\emph{Commun. {ACM}}} \bibinfo{volume}{12}, \bibinfo{number}{10} (\bibinfo{year}{1969}), \bibinfo{pages}{576--580}.
\newblock
\urldef\tempurl%
\url{https://doi.org/10.1145/363235.363259}
\showURL{%
\tempurl}


\bibitem[Hou et~al\mbox{.}(2022)]%
        {hou-etal-2022-metaprompting}
\bibfield{author}{\bibinfo{person}{Yutai Hou}, \bibinfo{person}{Hongyuan Dong}, \bibinfo{person}{Xinghao Wang}, \bibinfo{person}{Bohan Li}, {and} \bibinfo{person}{Wanxiang Che}.} \bibinfo{year}{2022}\natexlab{}.
\newblock \showarticletitle{MetaPrompting: Learning to Learn Better Prompts}. In \bibinfo{booktitle}{\emph{Proceedings of the 29th International Conference on Computational Linguistics, {COLING} 2022, Gyeongju, Republic of Korea, October 12-17, 2022}}, \bibfield{editor}{\bibinfo{person}{Nicoletta Calzolari}, \bibinfo{person}{Chu{-}Ren Huang}, \bibinfo{person}{Hansaem Kim}, \bibinfo{person}{James Pustejovsky}, \bibinfo{person}{Leo Wanner}, \bibinfo{person}{Key{-}Sun Choi}, \bibinfo{person}{Pum{-}Mo Ryu}, \bibinfo{person}{Hsin{-}Hsi Chen}, \bibinfo{person}{Lucia Donatelli}, \bibinfo{person}{Heng Ji}, \bibinfo{person}{Sadao Kurohashi}, \bibinfo{person}{Patrizia Paggio}, \bibinfo{person}{Nianwen Xue}, \bibinfo{person}{Seokhwan Kim}, \bibinfo{person}{Younggyun Hahm}, \bibinfo{person}{Zhong He}, \bibinfo{person}{Tony~Kyungil Lee}, \bibinfo{person}{Enrico Santus}, \bibinfo{person}{Francis Bond}, {and} \bibinfo{person}{Seung{-}Hoon Na}} (Eds.). \bibinfo{publisher}{International Committee on Computational
  Linguistics}, \bibinfo{pages}{3251--3262}.
\newblock
\urldef\tempurl%
\url{https://aclanthology.org/2022.coling-1.287}
\showURL{%
\tempurl}


\bibitem[Irfan et~al\mbox{.}(2022)]%
        {DBLP:conf/issta/IrfanPRRT22}
\bibfield{author}{\bibinfo{person}{Ahmed Irfan}, \bibinfo{person}{Sorawee Porncharoenwase}, \bibinfo{person}{Zvonimir Rakamaric}, \bibinfo{person}{Neha Rungta}, {and} \bibinfo{person}{Emina Torlak}.} \bibinfo{year}{2022}\natexlab{}.
\newblock \showarticletitle{Testing Dafny (experience paper)}. In \bibinfo{booktitle}{\emph{{ISSTA} '22: 31st {ACM} {SIGSOFT} International Symposium on Software Testing and Analysis, Virtual Event, South Korea, July 18 - 22, 2022}}, \bibfield{editor}{\bibinfo{person}{Sukyoung Ryu} {and} \bibinfo{person}{Yannis Smaragdakis}} (Eds.). \bibinfo{publisher}{{ACM}}, \bibinfo{pages}{556--567}.
\newblock
\urldef\tempurl%
\url{https://doi.org/10.1145/3533767.3534382}
\showURL{%
\tempurl}


\bibitem[Jacobs and Beurdouche(2022)]%
        {mozilla}
\bibfield{author}{\bibinfo{person}{Kevin Jacobs} {and} \bibinfo{person}{Benjamin Beurdouche}.} \bibinfo{year}{2022}\natexlab{}.
\newblock \bibinfo{title}{Performance Improvements via Formally-Verified Cryptography in Firefox}.
\newblock \bibinfo{howpublished}{\url{https://blog.mozilla.org/security/2020/07/06/performance-improvements-viaformally-verified-cryptography-in-firefox/}}.
\newblock
\newblock
\shownote{[Online], [Accessed: 2023-09-20]}.


\bibitem[Jiang et~al\mbox{.}(2022a)]%
        {DBLP:conf/nips/JiangLTCOMWJ22}
\bibfield{author}{\bibinfo{person}{Albert~Qiaochu Jiang}, \bibinfo{person}{Wenda Li}, \bibinfo{person}{Szymon Tworkowski}, \bibinfo{person}{Konrad Czechowski}, \bibinfo{person}{Tomasz Odrzyg{\'{o}}zdz}, \bibinfo{person}{Piotr Milos}, \bibinfo{person}{Yuhuai Wu}, {and} \bibinfo{person}{Mateja Jamnik}.} \bibinfo{year}{2022}\natexlab{a}.
\newblock \showarticletitle{Thor: Wielding Hammers to Integrate Language Models and Automated Theorem Provers}. In \bibinfo{booktitle}{\emph{Advances in Neural Information Processing Systems 35: Annual Conference on Neural Information Processing Systems 2022, NeurIPS 2022, New Orleans, LA, USA, November 28 - December 9, 2022}}, \bibfield{editor}{\bibinfo{person}{Sanmi Koyejo}, \bibinfo{person}{S.~Mohamed}, \bibinfo{person}{A.~Agarwal}, \bibinfo{person}{Danielle Belgrave}, \bibinfo{person}{K.~Cho}, {and} \bibinfo{person}{A.~Oh}} (Eds.).
\newblock
\urldef\tempurl%
\url{http://papers.nips.cc/paper\_files/paper/2022/hash/377c25312668e48f2e531e2f2c422483-Abstract-Conference.html}
\showURL{%
\tempurl}


\bibitem[Jiang et~al\mbox{.}(2022b)]%
        {jiang2022}
\bibfield{author}{\bibinfo{person}{Ellen Jiang}, \bibinfo{person}{Kristen Olson}, \bibinfo{person}{Edwin Toh}, \bibinfo{person}{Alejandra Molina}, \bibinfo{person}{Aaron Donsbach}, \bibinfo{person}{Michael Terry}, {and} \bibinfo{person}{Carrie~J. Cai}.} \bibinfo{year}{2022}\natexlab{b}.
\newblock \showarticletitle{PromptMaker: Prompt-based Prototyping with Large Language Models}. In \bibinfo{booktitle}{\emph{{CHI} '22: {CHI} Conference on Human Factors in Computing Systems, New Orleans, LA, USA, 29 April 2022 - 5 May 2022, Extended Abstracts}}, \bibfield{editor}{\bibinfo{person}{Simone D.~J. Barbosa}, \bibinfo{person}{Cliff Lampe}, \bibinfo{person}{Caroline Appert}, {and} \bibinfo{person}{David~A. Shamma}} (Eds.). \bibinfo{publisher}{{ACM}}, \bibinfo{pages}{35:1--35:8}.
\newblock
\urldef\tempurl%
\url{https://doi.org/10.1145/3491101.3503564}
\showDOI{\tempurl}


\bibitem[Jones and Misra(2021)]%
        {Jones2021}
\bibfield{author}{\bibinfo{person}{Cliff~B Jones} {and} \bibinfo{person}{Jayadev Misra}.} \bibinfo{year}{2021}\natexlab{}.
\newblock \bibinfo{booktitle}{\emph{Theories of Programming: The Life and Works of Tony Hoare}}.
\newblock \bibinfo{publisher}{Morgan \& Claypool}.
\newblock


\bibitem[Klein et~al\mbox{.}(2009)]%
        {DBLP:conf/sosp/KleinEHACDEEKNSTW09}
\bibfield{author}{\bibinfo{person}{Gerwin Klein}, \bibinfo{person}{Kevin Elphinstone}, \bibinfo{person}{Gernot Heiser}, \bibinfo{person}{June Andronick}, \bibinfo{person}{David~A. Cock}, \bibinfo{person}{Philip Derrin}, \bibinfo{person}{Dhammika Elkaduwe}, \bibinfo{person}{Kai Engelhardt}, \bibinfo{person}{Rafal Kolanski}, \bibinfo{person}{Michael Norrish}, \bibinfo{person}{Thomas Sewell}, \bibinfo{person}{Harvey Tuch}, {and} \bibinfo{person}{Simon Winwood}.} \bibinfo{year}{2009}\natexlab{}.
\newblock \showarticletitle{se{L}4: formal verification of an OS kernel.}. In \bibinfo{booktitle}{\emph{Proceedings of the 22nd {ACM} Symposium on Operating Systems Principles 2009, {SOSP} 2009, Big Sky, Montana, USA, October 11-14, 2009}}, \bibfield{editor}{\bibinfo{person}{Jeanna~Neefe Matthews} {and} \bibinfo{person}{Thomas~E. Anderson}} (Eds.). \bibinfo{publisher}{{ACM}}, \bibinfo{pages}{207--220}.
\newblock
\urldef\tempurl%
\url{https://doi.org/10.1145/1629575.1629596}
\showDOI{\tempurl}


\bibitem[Kulal et~al\mbox{.}(2019)]%
        {DBLP:conf/nips/KulalPC0PAL19}
\bibfield{author}{\bibinfo{person}{Sumith Kulal}, \bibinfo{person}{Panupong Pasupat}, \bibinfo{person}{Kartik Chandra}, \bibinfo{person}{Mina Lee}, \bibinfo{person}{Oded Padon}, \bibinfo{person}{Alex Aiken}, {and} \bibinfo{person}{Percy Liang}.} \bibinfo{year}{2019}\natexlab{}.
\newblock \showarticletitle{SPoC: Search-based Pseudocode to Code}. In \bibinfo{booktitle}{\emph{Advances in Neural Information Processing Systems 32: Annual Conference on Neural Information Processing Systems 2019, NeurIPS 2019, December 8-14, 2019, Vancouver, BC, Canada}}, \bibfield{editor}{\bibinfo{person}{Hanna~M. Wallach}, \bibinfo{person}{Hugo Larochelle}, \bibinfo{person}{Alina Beygelzimer}, \bibinfo{person}{Florence d'Alch{\'{e}}{-}Buc}, \bibinfo{person}{Emily~B. Fox}, {and} \bibinfo{person}{Roman Garnett}} (Eds.). \bibinfo{pages}{11883--11894}.
\newblock


\bibitem[Leino(2017)]%
        {DBLP:journals/software/Leino17}
\bibfield{author}{\bibinfo{person}{K.~Rustan~M. Leino}.} \bibinfo{year}{2017}\natexlab{}.
\newblock \showarticletitle{Accessible Software Verification with Dafny}.
\newblock \bibinfo{journal}{\emph{{IEEE} Softw.}} \bibinfo{volume}{34}, \bibinfo{number}{6} (\bibinfo{year}{2017}), \bibinfo{pages}{94--97}.
\newblock
\urldef\tempurl%
\url{https://doi.org/10.1109/MS.2017.4121212}
\showURL{%
\tempurl}


\bibitem[Leino(2023)]%
        {dafny2023}
\bibfield{author}{\bibinfo{person}{K.~Rustan~M. Leino}.} \bibinfo{year}{2023}\natexlab{}.
\newblock \bibinfo{booktitle}{\emph{Program Proofs}}.
\newblock \bibinfo{publisher}{{MIT} Press}.
\newblock


\bibitem[Leino and Nelson(1998)]%
        {rustan1998extended}
\bibfield{author}{\bibinfo{person}{K.~Rustan~M. Leino} {and} \bibinfo{person}{Greg Nelson}.} \bibinfo{year}{1998}\natexlab{}.
\newblock \showarticletitle{An extended static checker for {M}odula-3}. In \bibinfo{booktitle}{\emph{International Conference on Compiler Construction}}. Springer, \bibinfo{pages}{302--305}.
\newblock


\bibitem[Leroy(2009)]%
        {DBLP:journals/cacm/Leroy09}
\bibfield{author}{\bibinfo{person}{Xavier Leroy}.} \bibinfo{year}{2009}\natexlab{}.
\newblock \showarticletitle{Formal verification of a realistic compiler}.
\newblock \bibinfo{journal}{\emph{Commun. {ACM}}} \bibinfo{volume}{52}, \bibinfo{number}{7} (\bibinfo{year}{2009}), \bibinfo{pages}{107--115}.
\newblock
\urldef\tempurl%
\url{https://doi.org/10.1145/1538788.1538814}
\showDOI{\tempurl}


\bibitem[Lewis et~al\mbox{.}(2020)]%
        {RAG}
\bibfield{author}{\bibinfo{person}{Patrick S.~H. Lewis}, \bibinfo{person}{Ethan Perez}, \bibinfo{person}{Aleksandra Piktus}, \bibinfo{person}{Fabio Petroni}, \bibinfo{person}{Vladimir Karpukhin}, \bibinfo{person}{Naman Goyal}, \bibinfo{person}{Heinrich K{\"{u}}ttler}, \bibinfo{person}{Mike Lewis}, \bibinfo{person}{Wen{-}tau Yih}, \bibinfo{person}{Tim Rockt{\"{a}}schel}, \bibinfo{person}{Sebastian Riedel}, {and} \bibinfo{person}{Douwe Kiela}.} \bibinfo{year}{2020}\natexlab{}.
\newblock \showarticletitle{Retrieval-Augmented Generation for Knowledge-Intensive {NLP} Tasks}. In \bibinfo{booktitle}{\emph{Advances in Neural Information Processing Systems 33: Annual Conference on Neural Information Processing Systems 2020, NeurIPS 2020, December 6-12, 2020, virtual}}, \bibfield{editor}{\bibinfo{person}{Hugo Larochelle}, \bibinfo{person}{Marc'Aurelio Ranzato}, \bibinfo{person}{Raia Hadsell}, \bibinfo{person}{Maria{-}Florina Balcan}, {and} \bibinfo{person}{Hsuan{-}Tien Lin}} (Eds.).
\newblock
\urldef\tempurl%
\url{https://proceedings.neurips.cc/paper/2020/hash/6b493230205f780e1bc26945df7481e5-Abstract.html}
\showURL{%
\tempurl}


\bibitem[Li et~al\mbox{.}(2022a)]%
        {DBLP:conf/sigsoft/LiYJYLHLZ22}
\bibfield{author}{\bibinfo{person}{Lingwei Li}, \bibinfo{person}{Li Yang}, \bibinfo{person}{Huaxi Jiang}, \bibinfo{person}{Jun Yan}, \bibinfo{person}{Tiejian Luo}, \bibinfo{person}{Zihan Hua}, \bibinfo{person}{Geng Liang}, {and} \bibinfo{person}{Chun Zuo}.} \bibinfo{year}{2022}\natexlab{a}.
\newblock \showarticletitle{{AUGER:} automatically generating review comments with pre-training models}. In \bibinfo{booktitle}{\emph{{ESEC/FSE}}}, \bibfield{editor}{\bibinfo{person}{Abhik Roychoudhury}, \bibinfo{person}{Cristian Cadar}, {and} \bibinfo{person}{Miryung Kim}} (Eds.). \bibinfo{publisher}{{ACM}}, \bibinfo{pages}{1009--1021}.
\newblock
\urldef\tempurl%
\url{https://doi.org/10.1145/3540250.3549099}
\showURL{%
\tempurl}


\bibitem[Li et~al\mbox{.}(2022b)]%
        {DBLP:journals/corr/abs-2211-06411}
\bibfield{author}{\bibinfo{person}{Liyi Li}, \bibinfo{person}{Mingwei Zhu}, \bibinfo{person}{Yi Lee}, \bibinfo{person}{Le Chang}, {and} \bibinfo{person}{Xiaodi Wu}.} \bibinfo{year}{2022}\natexlab{b}.
\newblock \showarticletitle{Quantum Natural Proof: {A} New Perspective of Hybrid Quantum-Classical Program Verification}.
\newblock \bibinfo{journal}{\emph{CoRR}}  \bibinfo{volume}{abs/2211.06411} (\bibinfo{year}{2022}).
\newblock
\urldef\tempurl%
\url{https://doi.org/10.48550/arXiv.2211.06411}
\showURL{%
\tempurl}


\bibitem[Li et~al\mbox{.}(2023)]%
        {DBLP:conf/acl/2023/LiMaking}
\bibfield{author}{\bibinfo{person}{Yifei Li}, \bibinfo{person}{Zeqi Lin}, \bibinfo{person}{Shizhuo Zhang}, \bibinfo{person}{Qiang Fu}, \bibinfo{person}{Bei Chen}, \bibinfo{person}{Jian-Guang Lou}, {and} \bibinfo{person}{Weizhu Chen}.} \bibinfo{year}{2023}\natexlab{}.
\newblock \showarticletitle{Making Language Models Better Reasoners with Step-Aware Verifier}. In \bibinfo{booktitle}{\emph{Proceedings of the 61st Annual Meeting of the Association for Computational Linguistics (Volume 1: Long Papers)}}, \bibfield{editor}{\bibinfo{person}{Anna Rogers}, \bibinfo{person}{Jordan Boyd-Graber}, {and} \bibinfo{person}{Naoaki Okazaki}} (Eds.). \bibinfo{publisher}{Association for Computational Linguistics}, \bibinfo{address}{Toronto, Canada}, \bibinfo{pages}{5315--5333}.
\newblock
\urldef\tempurl%
\url{https://doi.org/10.18653/v1/2023.acl-long.291}
\showDOI{\tempurl}


\bibitem[Liu and Chilton(2022)]%
        {liu2022}
\bibfield{author}{\bibinfo{person}{Vivian Liu} {and} \bibinfo{person}{Lydia~B. Chilton}.} \bibinfo{year}{2022}\natexlab{}.
\newblock \showarticletitle{Design Guidelines for Prompt Engineering Text-to-Image Generative Models}. In \bibinfo{booktitle}{\emph{{CHI} '22: {CHI} Conference on Human Factors in Computing Systems, New Orleans, LA, USA, 29 April 2022 - 5 May 2022}}, \bibfield{editor}{\bibinfo{person}{Simone D.~J. Barbosa}, \bibinfo{person}{Cliff Lampe}, \bibinfo{person}{Caroline Appert}, \bibinfo{person}{David~A. Shamma}, \bibinfo{person}{Steven~Mark Drucker}, \bibinfo{person}{Julie~R. Williamson}, {and} \bibinfo{person}{Koji Yatani}} (Eds.). \bibinfo{publisher}{{ACM}}, \bibinfo{pages}{384:1--384:23}.
\newblock
\urldef\tempurl%
\url{https://doi.org/10.1145/3491102.3501825}
\showDOI{\tempurl}


\bibitem[LLVM(2023)]%
        {LLVM}
\bibfield{author}{\bibinfo{person}{LLVM}.} \bibinfo{year}{2023}\natexlab{}.
\newblock \bibinfo{title}{The LLVM Compiler Infrastructure}.
\newblock \bibinfo{howpublished}{\url{https://llvm.org/}}.
\newblock
\newblock
\shownote{[Online], [Accessed: 2023-09-20]}.


\bibitem[Madaan et~al\mbox{.}(2022)]%
        {DBLP:conf/emnlp/MadaanZ0YN22}
\bibfield{author}{\bibinfo{person}{Aman Madaan}, \bibinfo{person}{Shuyan Zhou}, \bibinfo{person}{Uri Alon}, \bibinfo{person}{Yiming Yang}, {and} \bibinfo{person}{Graham Neubig}.} \bibinfo{year}{2022}\natexlab{}.
\newblock \showarticletitle{Language Models of Code are Few-Shot Commonsense Learners}. In \bibinfo{booktitle}{\emph{{EMNLP}}}, \bibfield{editor}{\bibinfo{person}{Yoav Goldberg}, \bibinfo{person}{Zornitsa Kozareva}, {and} \bibinfo{person}{Yue Zhang}} (Eds.). \bibinfo{publisher}{Association for Computational Linguistics}, \bibinfo{pages}{1384--1403}.
\newblock


\bibitem[{Mark Chen et al.}(2021)]%
        {DBLP:journals/corr/abs-2107-03374}
\bibfield{author}{\bibinfo{person}{{Mark Chen et al.}}} \bibinfo{year}{2021}\natexlab{}.
\newblock \showarticletitle{Evaluating Large Language Models Trained on Code}.
\newblock \bibinfo{journal}{\emph{CoRR}}  \bibinfo{volume}{abs/2107.03374} (\bibinfo{year}{2021}).
\newblock
\urldef\tempurl%
\url{https://arxiv.org/abs/2107.03374}
\showURL{%
\tempurl}


\bibitem[Mashhadi and Hemmati(2021)]%
        {DBLP:conf/msr/MashhadiH21}
\bibfield{author}{\bibinfo{person}{Ehsan Mashhadi} {and} \bibinfo{person}{Hadi Hemmati}.} \bibinfo{year}{2021}\natexlab{}.
\newblock \showarticletitle{Applying CodeBERT for Automated Program Repair of Java Simple Bugs}. In \bibinfo{booktitle}{\emph{{MSR}}}. \bibinfo{pages}{505--509}.
\newblock
\urldef\tempurl%
\url{https://doi.org/10.1109/MSR52588.2021.00063}
\showURL{%
\tempurl}


\bibitem[Mastropaolo et~al\mbox{.}(2023)]%
        {DBLP:journals/tse/MastropaoloCNSPOB23}
\bibfield{author}{\bibinfo{person}{Antonio Mastropaolo}, \bibinfo{person}{Nathan Cooper}, \bibinfo{person}{David Nader{-}Palacio}, \bibinfo{person}{Simone Scalabrino}, \bibinfo{person}{Denys Poshyvanyk}, \bibinfo{person}{Rocco Oliveto}, {and} \bibinfo{person}{Gabriele Bavota}.} \bibinfo{year}{2023}\natexlab{}.
\newblock \showarticletitle{Using Transfer Learning for Code-Related Tasks}.
\newblock \bibinfo{journal}{\emph{{IEEE} Trans. Software Eng.}} \bibinfo{volume}{49}, \bibinfo{number}{4} (\bibinfo{year}{2023}), \bibinfo{pages}{1580--1598}.
\newblock
\urldef\tempurl%
\url{https://doi.org/10.1109/TSE.2022.3183297}
\showDOI{\tempurl}


\bibitem[McKenna et~al\mbox{.}(2023)]%
        {mckenna2023sources}
\bibfield{author}{\bibinfo{person}{Nick McKenna}, \bibinfo{person}{Tianyi Li}, \bibinfo{person}{Liang Cheng}, \bibinfo{person}{Mohammad~Javad Hosseini}, \bibinfo{person}{Mark Johnson}, {and} \bibinfo{person}{Mark Steedman}.} \bibinfo{year}{2023}\natexlab{}.
\newblock \showarticletitle{Sources of Hallucination by Large Language Models on Inference Tasks}. In \bibinfo{booktitle}{\emph{Findings of the Association for Computational Linguistics: {EMNLP} 2023, Singapore, December 6-10, 2023}}, \bibfield{editor}{\bibinfo{person}{Houda Bouamor}, \bibinfo{person}{Juan Pino}, {and} \bibinfo{person}{Kalika Bali}} (Eds.). \bibinfo{publisher}{Association for Computational Linguistics}, \bibinfo{pages}{2758--2774}.
\newblock
\urldef\tempurl%
\url{https://aclanthology.org/2023.findings-emnlp.182}
\showURL{%
\tempurl}


\bibitem[Meadows and Freitas(2022)]%
        {DBLP:journals/corr/abs-2205-15231}
\bibfield{author}{\bibinfo{person}{Jordan Meadows} {and} \bibinfo{person}{Andr{\'{e}} Freitas}.} \bibinfo{year}{2022}\natexlab{}.
\newblock \showarticletitle{A Survey in Mathematical Language Processing}.
\newblock \bibinfo{journal}{\emph{CoRR}}  \bibinfo{volume}{abs/2205.15231} (\bibinfo{year}{2022}).
\newblock
\urldef\tempurl%
\url{https://doi.org/10.48550/arXiv.2205.15231}
\showURL{%
\tempurl}


\bibitem[Meyer(1992)]%
        {meyer1992applying}
\bibfield{author}{\bibinfo{person}{Bertrand Meyer}.} \bibinfo{year}{1992}\natexlab{}.
\newblock \showarticletitle{Applying 'Design by Contract'}.
\newblock \bibinfo{journal}{\emph{Computer}} \bibinfo{volume}{25}, \bibinfo{number}{10} (\bibinfo{year}{1992}), \bibinfo{pages}{40--51}.
\newblock


\bibitem[Microsoft(2023a)]%
        {dafnytsite}
\bibfield{author}{\bibinfo{person}{Microsoft}.} \bibinfo{year}{2023}\natexlab{a}.
\newblock \bibinfo{title}{The Dafny Programming and Verification Language}.
\newblock \bibinfo{howpublished}{\url{https://dafny.org/}}.
\newblock
\newblock
\shownote{[Online], [Accessed: 2023-09-20]}.


\bibitem[Microsoft(2023b)]%
        {microsoftresearch}
\bibfield{author}{\bibinfo{person}{Microsoft}.} \bibinfo{year}{2023}\natexlab{b}.
\newblock \bibinfo{title}{Microsoft Research}.
\newblock \bibinfo{howpublished}{\url{https://www.microsoft.com/en-us/research/}}.
\newblock
\newblock
\shownote{[Online], [Accessed: 2023-09-20]}.


\bibitem[Murray et~al\mbox{.}(2013)]%
        {DBLP:conf/sp/MurrayMBGBSLGK13}
\bibfield{author}{\bibinfo{person}{Toby~C. Murray}, \bibinfo{person}{Daniel Matichuk}, \bibinfo{person}{Matthew Brassil}, \bibinfo{person}{Peter Gammie}, \bibinfo{person}{Timothy Bourke}, \bibinfo{person}{Sean Seefried}, \bibinfo{person}{Corey Lewis}, \bibinfo{person}{Xin Gao}, {and} \bibinfo{person}{Gerwin Klein}.} \bibinfo{year}{2013}\natexlab{}.
\newblock \showarticletitle{seL4: From General Purpose to a Proof of Information Flow Enforcement}. In \bibinfo{booktitle}{\emph{2013 {IEEE} Symposium on Security and Privacy, {SP} 2013, Berkeley, CA, USA, May 19-22, 2013}}. \bibinfo{publisher}{{IEEE} Computer Society}, \bibinfo{pages}{415--429}.
\newblock
\urldef\tempurl%
\url{https://doi.org/10.1109/SP.2013.35}
\showDOI{\tempurl}


\bibitem[Nashid et~al\mbox{.}(2023)]%
        {nashid2023retrieval}
\bibfield{author}{\bibinfo{person}{Noor Nashid}, \bibinfo{person}{Mifta Sintaha}, {and} \bibinfo{person}{Ali Mesbah}.} \bibinfo{year}{2023}\natexlab{}.
\newblock \showarticletitle{Retrieval-Based Prompt Selection for Code-Related Few-Shot Learning}. In \bibinfo{booktitle}{\emph{45th {IEEE/ACM} International Conference on Software Engineering, {ICSE} 2023, Melbourne, Australia, May 14-20, 2023}}. \bibinfo{publisher}{{IEEE}}, \bibinfo{pages}{2450--2462}.
\newblock
\urldef\tempurl%
\url{https://doi.org/10.1109/ICSE48619.2023.00205}
\showDOI{\tempurl}


\bibitem[Nederpelt et~al\mbox{.}(1994)]%
        {Automath}
\bibfield{editor}{\bibinfo{person}{R.P. Nederpelt}, \bibinfo{person}{J.H. Geuvers}, {and} \bibinfo{person}{R.C. de Vrijer}} (Eds.). \bibinfo{year}{1994}\natexlab{}.
\newblock \bibinfo{booktitle}{\emph{Selected Papers on Automath}}.
\newblock \bibinfo{publisher}{North Holland}.
\newblock


\bibitem[Noble(2024)]%
        {LearnEm}
\bibfield{author}{\bibinfo{person}{James Noble}.} \bibinfo{year}{2024}\natexlab{}.
\newblock \showarticletitle{Learn 'em {D}afny}. In \bibinfo{booktitle}{\emph{Dafny Workshop at {POPL}}} (London, England).
\newblock
\urldef\tempurl%
\url{https://popl24.sigplan.org/details/dafny-2024-papers/11/Learn-em-Dafny}
\showURL{%
\tempurl}


\bibitem[Noble et~al\mbox{.}(2022)]%
        {MPTP}
\bibfield{author}{\bibinfo{person}{James Noble}, \bibinfo{person}{David Streader}, \bibinfo{person}{Isaac~Oscar Gariano}, {and} \bibinfo{person}{Miniruwani Samarakoon}.} \bibinfo{year}{2022}\natexlab{}.
\newblock \showarticletitle{More Programming Than Programming: Teaching Formal Methods in a Software Engineering Programme}. In \bibinfo{booktitle}{\emph{{NASA} Formal Methods - 14th International Symposium, {NFM} 2022, Pasadena, CA, USA, May 24-27, 2022, Proceedings}} \emph{(\bibinfo{series}{Lecture Notes in Computer Science}, Vol.~\bibinfo{volume}{13260})}, \bibfield{editor}{\bibinfo{person}{Jyotirmoy~V. Deshmukh}, \bibinfo{person}{Klaus Havelund}, {and} \bibinfo{person}{Ivan Perez}} (Eds.). \bibinfo{publisher}{Springer}, \bibinfo{pages}{431--450}.
\newblock
\urldef\tempurl%
\url{https://doi.org/10.1007/978-3-031-06773-0\_23}
\showDOI{\tempurl}


\bibitem[OpenAI(2023a)]%
        {DBLP:journals/corr/abs-2303-08774}
\bibfield{author}{\bibinfo{person}{OpenAI}.} \bibinfo{year}{2023}\natexlab{a}.
\newblock \showarticletitle{{GPT-4} Technical Report}.
\newblock \bibinfo{journal}{\emph{CoRR}}  \bibinfo{volume}{abs/2303.08774} (\bibinfo{year}{2023}).
\newblock
\urldef\tempurl%
\url{https://doi.org/10.48550/arXiv.2303.08774}
\showURL{%
\tempurl}


\bibitem[OpenAI(2023b)]%
        {openai}
\bibfield{author}{\bibinfo{person}{OpenAI}.} \bibinfo{year}{2023}\natexlab{b}.
\newblock \bibinfo{title}{OpenAI}.
\newblock \bibinfo{howpublished}{\url{https://platform.openai.com/docs/models}}.
\newblock
\newblock
\shownote{[Online], [Accessed: 2023-09-20]}.


\bibitem[Qiao et~al\mbox{.}(2023)]%
        {DBLP:conf/acl/QiaoO0CYDTHC23}
\bibfield{author}{\bibinfo{person}{Shuofei Qiao}, \bibinfo{person}{Yixin Ou}, \bibinfo{person}{Ningyu Zhang}, \bibinfo{person}{Xiang Chen}, \bibinfo{person}{Yunzhi Yao}, \bibinfo{person}{Shumin Deng}, \bibinfo{person}{Chuanqi Tan}, \bibinfo{person}{Fei Huang}, {and} \bibinfo{person}{Huajun Chen}.} \bibinfo{year}{2023}\natexlab{}.
\newblock \showarticletitle{Reasoning with Language Model Prompting: {A} Survey}. In \bibinfo{booktitle}{\emph{Proceedings of the 61st Annual Meeting of the Association for Computational Linguistics (Volume 1: Long Papers), {ACL} 2023, Toronto, Canada, July 9-14, 2023}}, \bibfield{editor}{\bibinfo{person}{Anna Rogers}, \bibinfo{person}{Jordan~L. Boyd{-}Graber}, {and} \bibinfo{person}{Naoaki Okazaki}} (Eds.). \bibinfo{publisher}{Association for Computational Linguistics}, \bibinfo{pages}{5368--5393}.
\newblock
\urldef\tempurl%
\url{https://doi.org/10.18653/V1/2023.ACL-LONG.294}
\showDOI{\tempurl}


\bibitem[Radford et~al\mbox{.}(2018)]%
        {radford2018improving}
\bibfield{author}{\bibinfo{person}{Alec Radford}, \bibinfo{person}{Karthik Narasimhan}, \bibinfo{person}{Tim Salimans}, \bibinfo{person}{Ilya Sutskever}, {et~al\mbox{.}}} \bibinfo{year}{2018}\natexlab{}.
\newblock \bibinfo{title}{Improving language understanding by generative pre-training}.
\newblock \bibinfo{howpublished}{{OpenAI.com}}.
\newblock


\bibitem[Reynolds and McDonell(2021)]%
        {promptgramming2021}
\bibfield{author}{\bibinfo{person}{Laria Reynolds} {and} \bibinfo{person}{Kyle McDonell}.} \bibinfo{year}{2021}\natexlab{}.
\newblock \showarticletitle{Prompt Programming for Large Language Models: Beyond the Few-Shot Paradigm}. In \bibinfo{booktitle}{\emph{{CHI} '21: {CHI} Conference on Human Factors in Computing Systems, Virtual Event / Yokohama Japan, May 8-13, 2021, Extended Abstracts}}, \bibfield{editor}{\bibinfo{person}{Yoshifumi Kitamura}, \bibinfo{person}{Aaron Quigley}, \bibinfo{person}{Katherine Isbister}, {and} \bibinfo{person}{Takeo Igarashi}} (Eds.). \bibinfo{publisher}{{ACM}}, \bibinfo{pages}{314:1--314:7}.
\newblock
\urldef\tempurl%
\url{https://doi.org/10.1145/3411763.3451760}
\showDOI{\tempurl}


\bibitem[Ringer et~al\mbox{.}(2019)]%
        {FTPL-formalism}
\bibfield{author}{\bibinfo{person}{Talia Ringer}, \bibinfo{person}{Karl Palmskog}, \bibinfo{person}{Ilya Sergey}, \bibinfo{person}{Milos Gligoric}, {and} \bibinfo{person}{Zachary Tatlock}.} \bibinfo{year}{2019}\natexlab{}.
\newblock \showarticletitle{QED at Large: A Survey of Engineering of Formally Verified Software}.
\newblock \bibinfo{journal}{\emph{{Founds.\ Trends.\ Prog.\ Lang.}}} \bibinfo{volume}{5}, \bibinfo{number}{2-3} (\bibinfo{year}{2019}), \bibinfo{pages}{102--281}.
\newblock
\urldef\tempurl%
\url{https://doi.org/10.1561/2500000045}
\showDOI{\tempurl}


\bibitem[{Rohan Anil et al.}(2023)]%
        {DBLP:journals/corr/abs-2305-10403}
\bibfield{author}{\bibinfo{person}{{Rohan Anil et al.}}} \bibinfo{year}{2023}\natexlab{}.
\newblock \showarticletitle{PaLM 2 Technical Report}.
\newblock \bibinfo{journal}{\emph{CoRR}}  \bibinfo{volume}{abs/2305.10403} (\bibinfo{year}{2023}).
\newblock


\bibitem[Rozi{\`e}re et~al\mbox{.}(2023)]%
        {roziere2023code}
\bibfield{author}{\bibinfo{person}{Baptiste Rozi{\`e}re}, \bibinfo{person}{Jonas Gehring}, \bibinfo{person}{Fabian Gloeckle}, \bibinfo{person}{Sten Sootla}, \bibinfo{person}{Itai Gat}, \bibinfo{person}{Xiaoqing~Ellen Tan}, \bibinfo{person}{Yossi Adi}, \bibinfo{person}{Jingyu Liu}, \bibinfo{person}{Tal Remez}, \bibinfo{person}{J{\'e}r{\'e}my Rapin}, {et~al\mbox{.}}} \bibinfo{year}{2023}\natexlab{}.
\newblock \showarticletitle{Code llama: Open foundation models for code}.
\newblock \bibinfo{journal}{\emph{arXiv preprint arXiv:2308.12950}} (\bibinfo{year}{2023}).
\newblock


\bibitem[Salvagno and Taccone(2023)]%
        {hallucination2023}
\bibfield{author}{\bibinfo{person}{M. Salvagno} {and} \bibinfo{person}{A.G. Taccone, F.S.and~Gerli}.} \bibinfo{year}{2023}\natexlab{}.
\newblock \showarticletitle{Artificial intelligence hallucinations}.
\newblock \bibinfo{journal}{\emph{Crit Care}} (\bibinfo{year}{2023}).
\newblock
\urldef\tempurl%
\url{https://doi.org/10.1186/s13054-023-04473-y}
\showDOI{\tempurl}


\bibitem[Sanchez{-}Stern et~al\mbox{.}(2020)]%
        {DBLP:conf/pldi/Sanchez-SternAS20}
\bibfield{author}{\bibinfo{person}{Alex Sanchez{-}Stern}, \bibinfo{person}{Yousef Alhessi}, \bibinfo{person}{Lawrence~K. Saul}, {and} \bibinfo{person}{Sorin Lerner}.} \bibinfo{year}{2020}\natexlab{}.
\newblock \showarticletitle{Generating correctness proofs with neural networks}. In \bibinfo{booktitle}{\emph{Proceedings of the 4th {ACM} {SIGPLAN} International Workshop on Machine Learning and Programming Languages, MAPL@PLDI 2020, London, UK, June 15, 2020}}, \bibfield{editor}{\bibinfo{person}{Koushik Sen} {and} \bibinfo{person}{Mayur Naik}} (Eds.). \bibinfo{publisher}{{ACM}}, \bibinfo{pages}{1--10}.
\newblock
\urldef\tempurl%
\url{https://doi.org/10.1145/3394450.3397466}
\showDOI{\tempurl}


\bibitem[Sanchez{-}Stern et~al\mbox{.}(2023)]%
        {DBLP:journals/toplas/SanchezSternFZKBR23}
\bibfield{author}{\bibinfo{person}{Alex Sanchez{-}Stern}, \bibinfo{person}{Emily First}, \bibinfo{person}{Timothy Zhou}, \bibinfo{person}{Zhanna Kaufman}, \bibinfo{person}{Yuriy Brun}, {and} \bibinfo{person}{Talia Ringer}.} \bibinfo{year}{2023}\natexlab{}.
\newblock \showarticletitle{Passport: Improving Automated Formal Verification Using Identifiers}.
\newblock \bibinfo{journal}{\emph{{ACM} Trans. Program. Lang. Syst.}} \bibinfo{volume}{45}, \bibinfo{number}{2} (\bibinfo{year}{2023}), \bibinfo{pages}{12:1--12:30}.
\newblock
\urldef\tempurl%
\url{https://doi.org/10.1145/3593374}
\showDOI{\tempurl}


\bibitem[seL4 Project(2023)]%
        {seL4}
\bibfield{author}{\bibinfo{person}{seL4 Project}.} \bibinfo{year}{2023}\natexlab{}.
\newblock \bibinfo{title}{The seL4® Microkernel}.
\newblock \bibinfo{howpublished}{\url{https://sel4.systems/}}.
\newblock
\newblock
\shownote{[Online], [Accessed: 2023-09-20]}.


\bibitem[Siegmund et~al\mbox{.}(2015)]%
        {siegmunds}
\bibfield{author}{\bibinfo{person}{Janet Siegmund}, \bibinfo{person}{Norbert Siegmund}, {and} \bibinfo{person}{Sven Apel}.} \bibinfo{year}{2015}\natexlab{}.
\newblock \showarticletitle{Views on Internal and External Validity in Empirical Software Engineering}. In \bibinfo{booktitle}{\emph{37th {IEEE/ACM} International Conference on Software Engineering, {ICSE} 2015, Florence, Italy, May 16-24, 2015, Volume 1}}, \bibfield{editor}{\bibinfo{person}{Antonia Bertolino}, \bibinfo{person}{Gerardo Canfora}, {and} \bibinfo{person}{Sebastian~G. Elbaum}} (Eds.). \bibinfo{publisher}{{IEEE} Computer Society}, \bibinfo{pages}{9--19}.
\newblock
\urldef\tempurl%
\url{https://doi.org/10.1109/ICSE.2015.24}
\showDOI{\tempurl}


\bibitem[Sintaha et~al\mbox{.}(2023)]%
        {DBLP:journals/corr/abs-2205-00180}
\bibfield{author}{\bibinfo{person}{Mifta Sintaha}, \bibinfo{person}{Noor Nashid}, {and} \bibinfo{person}{Ali Mesbah}.} \bibinfo{year}{2023}\natexlab{}.
\newblock \showarticletitle{Katana: Dual Slicing Based Context for Learning Bug Fixes}.
\newblock \bibinfo{journal}{\emph{{ACM} Trans. Softw. Eng. Methodol.}} \bibinfo{volume}{32}, \bibinfo{number}{4} (\bibinfo{year}{2023}), \bibinfo{pages}{100:1--100:27}.
\newblock
\urldef\tempurl%
\url{https://doi.org/10.1145/3579640}
\showURL{%
\tempurl}


\bibitem[Tufano et~al\mbox{.}(2022a)]%
        {tufano2020unit}
\bibfield{author}{\bibinfo{person}{Michele Tufano}, \bibinfo{person}{Shao~Kun Deng}, \bibinfo{person}{Neel Sundaresan}, {and} \bibinfo{person}{Alexey Svyatkovskiy}.} \bibinfo{year}{2022}\natexlab{a}.
\newblock \showarticletitle{{METHODS2TEST:} {A} dataset of focal methods mapped to test cases}. In \bibinfo{booktitle}{\emph{19th {IEEE/ACM} International Conference on Mining Software Repositories, {MSR} 2022, Pittsburgh, PA, USA, May 23-24, 2022}}. \bibinfo{publisher}{{ACM}}, \bibinfo{pages}{299--303}.
\newblock
\urldef\tempurl%
\url{https://doi.org/10.1145/3524842.3528009}
\showDOI{\tempurl}


\bibitem[Tufano et~al\mbox{.}(2019)]%
        {DBLP:journals/tosem/TufanoWBPWP19}
\bibfield{author}{\bibinfo{person}{Michele Tufano}, \bibinfo{person}{Cody Watson}, \bibinfo{person}{Gabriele Bavota}, \bibinfo{person}{Massimiliano~Di Penta}, \bibinfo{person}{Martin White}, {and} \bibinfo{person}{Denys Poshyvanyk}.} \bibinfo{year}{2019}\natexlab{}.
\newblock \showarticletitle{An Empirical Study on Learning Bug-Fixing Patches in the Wild via Neural Machine Translation}.
\newblock \bibinfo{journal}{\emph{{ACM} Trans. Softw. Eng. Methodol.}} \bibinfo{volume}{28}, \bibinfo{number}{4} (\bibinfo{year}{2019}), \bibinfo{pages}{19:1--19:29}.
\newblock
\urldef\tempurl%
\url{https://doi.org/10.1145/3340544}
\showURL{%
\tempurl}


\bibitem[Tufano et~al\mbox{.}(2022b)]%
        {DBLP:conf/icse/TufanoMMPPB22}
\bibfield{author}{\bibinfo{person}{Rosalia Tufano}, \bibinfo{person}{Simone Masiero}, \bibinfo{person}{Antonio Mastropaolo}, \bibinfo{person}{Luca Pascarella}, \bibinfo{person}{Denys Poshyvanyk}, {and} \bibinfo{person}{Gabriele Bavota}.} \bibinfo{year}{2022}\natexlab{b}.
\newblock \showarticletitle{Using Pre-Trained Models to Boost Code Review Automation}. In \bibinfo{booktitle}{\emph{44th {IEEE/ACM} 44th International Conference on Software Engineering, {ICSE} 2022, Pittsburgh, PA, USA, May 25-27, 2022}}. \bibinfo{publisher}{{ACM}}, \bibinfo{pages}{2291--2302}.
\newblock
\urldef\tempurl%
\url{https://doi.org/10.1145/3510003.3510621}
\showURL{%
\tempurl}


\bibitem[Tufano et~al\mbox{.}(2023)]%
        {DBLP:conf/icse/TufanoPB23}
\bibfield{author}{\bibinfo{person}{Rosalia Tufano}, \bibinfo{person}{Luca Pascarella}, {and} \bibinfo{person}{Gabriele Bavota}.} \bibinfo{year}{2023}\natexlab{}.
\newblock \showarticletitle{Automating Code-Related Tasks Through Transformers: The Impact of Pre-training}. In \bibinfo{booktitle}{\emph{45th {IEEE/ACM} International Conference on Software Engineering, {ICSE} 2023, Melbourne, Australia, May 14-20, 2023}}. \bibinfo{publisher}{{IEEE}}, \bibinfo{pages}{2425--2437}.
\newblock
\urldef\tempurl%
\url{https://doi.org/10.1109/ICSE48619.2023.00203}
\showDOI{\tempurl}


\bibitem[Viera et~al\mbox{.}(2005)]%
        {viera2005understanding}
\bibfield{author}{\bibinfo{person}{Anthony~J Viera}, \bibinfo{person}{Joanne~M Garrett}, {et~al\mbox{.}}} \bibinfo{year}{2005}\natexlab{}.
\newblock \showarticletitle{Understanding interobserver agreement: the kappa statistic}.
\newblock \bibinfo{journal}{\emph{Fam med}} \bibinfo{volume}{37}, \bibinfo{number}{5} (\bibinfo{year}{2005}), \bibinfo{pages}{360--363}.
\newblock
\urldef\tempurl%
\url{https://pubmed.ncbi.nlm.nih.gov/15883903/}
\showURL{%
\tempurl}


\bibitem[Wang et~al\mbox{.}(2023)]%
        {DBLP:journals/corr/abs-2305-07922}
\bibfield{author}{\bibinfo{person}{Yue Wang}, \bibinfo{person}{Hung Le}, \bibinfo{person}{Akhilesh Gotmare}, \bibinfo{person}{Nghi D.~Q. Bui}, \bibinfo{person}{Junnan Li}, {and} \bibinfo{person}{Steven C.~H. Hoi}.} \bibinfo{year}{2023}\natexlab{}.
\newblock \showarticletitle{CodeT5+: Open Code Large Language Models for Code Understanding and Generation}.
\newblock  (\bibinfo{year}{2023}), \bibinfo{pages}{1069--1088}.
\newblock
\urldef\tempurl%
\url{https://aclanthology.org/2023.emnlp-main.68}
\showURL{%
\tempurl}


\bibitem[Wang et~al\mbox{.}(2021)]%
        {DBLP:conf/emnlp/0034WJH21}
\bibfield{author}{\bibinfo{person}{Y. Wang}, \bibinfo{person}{W. Wang}, \bibinfo{person}{S.R. Joty}, {and} \bibinfo{person}{S.C.H. Hoi}.} \bibinfo{year}{2021}\natexlab{}.
\newblock \showarticletitle{CodeT5: Identifier-aware Unified Pre-trained Encoder-Decoder Models for Code Understanding and Generation}. In \bibinfo{booktitle}{\emph{{EMNLP}}}.
\newblock
\urldef\tempurl%
\url{https://doi.org/10.18653/v1/2021.emnlp-main.685}
\showDOI{\tempurl}


\bibitem[Watson et~al\mbox{.}(2022)]%
        {DBLP:journals/tosem/WatsonCNMP22}
\bibfield{author}{\bibinfo{person}{Cody Watson}, \bibinfo{person}{Nathan Cooper}, \bibinfo{person}{David Nader{-}Palacio}, \bibinfo{person}{Kevin Moran}, {and} \bibinfo{person}{Denys Poshyvanyk}.} \bibinfo{year}{2022}\natexlab{}.
\newblock \showarticletitle{A Systematic Literature Review on the Use of Deep Learning in Software Engineering Research}.
\newblock \bibinfo{journal}{\emph{{ACM} Trans. Softw. Eng. Methodol.}} \bibinfo{volume}{31}, \bibinfo{number}{2} (\bibinfo{year}{2022}), \bibinfo{pages}{32:1--32:58}.
\newblock
\urldef\tempurl%
\url{https://doi.org/10.1145/3485275}
\showURL{%
\tempurl}


\bibitem[Wayne(2018)]%
        {wayne2018temporal}
\bibfield{author}{\bibinfo{person}{Hillel Wayne}.} \bibinfo{year}{2018}\natexlab{}.
\newblock \showarticletitle{Temporal Logic}.
\newblock In \bibinfo{booktitle}{\emph{Practical TLA+}}. \bibinfo{publisher}{Springer}, \bibinfo{pages}{97--110}.
\newblock


\bibitem[Wei et~al\mbox{.}(2022)]%
        {wei2022chain}
\bibfield{author}{\bibinfo{person}{Jason Wei}, \bibinfo{person}{Xuezhi Wang}, \bibinfo{person}{Dale Schuurmans}, \bibinfo{person}{Maarten Bosma}, \bibinfo{person}{Brian Ichter}, \bibinfo{person}{Fei Xia}, \bibinfo{person}{Ed~H. Chi}, \bibinfo{person}{Quoc~V. Le}, {and} \bibinfo{person}{Denny Zhou}.} \bibinfo{year}{2022}\natexlab{}.
\newblock \showarticletitle{Chain-of-Thought Prompting Elicits Reasoning in Large Language Models}. In \bibinfo{booktitle}{\emph{Advances in Neural Information Processing Systems 35: Annual Conference on Neural Information Processing Systems 2022, NeurIPS 2022, New Orleans, LA, USA, November 28 - December 9, 2022}}, \bibfield{editor}{\bibinfo{person}{Sanmi Koyejo}, \bibinfo{person}{S.~Mohamed}, \bibinfo{person}{A.~Agarwal}, \bibinfo{person}{Danielle Belgrave}, \bibinfo{person}{K.~Cho}, {and} \bibinfo{person}{A.~Oh}} (Eds.).
\newblock
\urldef\tempurl%
\url{http://papers.nips.cc/paper\_files/paper/2022/hash/9d5609613524ecf4f15af0f7b31abca4-Abstract-Conference.html}
\showURL{%
\tempurl}


\bibitem[Wu et~al\mbox{.}(2022)]%
        {DBLP:conf/nips/WuJLRSJS22}
\bibfield{author}{\bibinfo{person}{Yuhuai Wu}, \bibinfo{person}{Albert~Qiaochu Jiang}, \bibinfo{person}{Wenda Li}, \bibinfo{person}{Markus~N. Rabe}, \bibinfo{person}{Charles Staats}, \bibinfo{person}{Mateja Jamnik}, {and} \bibinfo{person}{Christian Szegedy}.} \bibinfo{year}{2022}\natexlab{}.
\newblock \showarticletitle{Autoformalization with Large Language Models}. In \bibinfo{booktitle}{\emph{Advances in Neural Information Processing Systems 35: Annual Conference on Neural Information Processing Systems 2022, NeurIPS 2022, New Orleans, LA, USA, November 28 - December 9, 2022}}, \bibfield{editor}{\bibinfo{person}{Sanmi Koyejo}, \bibinfo{person}{S.~Mohamed}, \bibinfo{person}{A.~Agarwal}, \bibinfo{person}{Danielle Belgrave}, \bibinfo{person}{K.~Cho}, {and} \bibinfo{person}{A.~Oh}} (Eds.).
\newblock
\urldef\tempurl%
\url{http://papers.nips.cc/paper\_files/paper/2022/hash/d0c6bc641a56bebee9d985b937307367-Abstract-Conference.html}
\showURL{%
\tempurl}


\bibitem[Yang and Deng(2019)]%
        {DBLP:conf/icml/YangD19}
\bibfield{author}{\bibinfo{person}{Kaiyu Yang} {and} \bibinfo{person}{Jia Deng}.} \bibinfo{year}{2019}\natexlab{}.
\newblock \showarticletitle{Learning to Prove Theorems via Interacting with Proof Assistants}. In \bibinfo{booktitle}{\emph{Proceedings of the 36th International Conference on Machine Learning, {ICML} 2019, 9-15 June 2019, Long Beach, California, {USA}}} \emph{(\bibinfo{series}{Proceedings of Machine Learning Research}, Vol.~\bibinfo{volume}{97})}, \bibfield{editor}{\bibinfo{person}{Kamalika Chaudhuri} {and} \bibinfo{person}{Ruslan Salakhutdinov}} (Eds.). \bibinfo{publisher}{{PMLR}}, \bibinfo{pages}{6984--6994}.
\newblock
\urldef\tempurl%
\url{http://proceedings.mlr.press/v97/yang19a.html}
\showURL{%
\tempurl}


\bibitem[Yang et~al\mbox{.}(2011)]%
        {DBLP:conf/pldi/YangCER11}
\bibfield{author}{\bibinfo{person}{Xuejun Yang}, \bibinfo{person}{Yang Chen}, \bibinfo{person}{Eric Eide}, {and} \bibinfo{person}{John Regehr}.} \bibinfo{year}{2011}\natexlab{}.
\newblock \showarticletitle{Finding and understanding bugs in {C} compilers}. In \bibinfo{booktitle}{\emph{Proceedings of the 32nd {ACM} {SIGPLAN} Conference on Programming Language Design and Implementation, {PLDI} 2011, San Jose, CA, USA, June 4-8, 2011}}, \bibfield{editor}{\bibinfo{person}{Mary~W. Hall} {and} \bibinfo{person}{David~A. Padua}} (Eds.). \bibinfo{publisher}{{ACM}}, \bibinfo{pages}{283--294}.
\newblock
\urldef\tempurl%
\url{https://doi.org/10.1145/1993498.1993532}
\showDOI{\tempurl}


\bibitem[Yang et~al\mbox{.}(2022)]%
        {DBLP:journals/csur/YangXLG22}
\bibfield{author}{\bibinfo{person}{Yanming Yang}, \bibinfo{person}{Xin Xia}, \bibinfo{person}{David Lo}, {and} \bibinfo{person}{John~C. Grundy}.} \bibinfo{year}{2022}\natexlab{}.
\newblock \showarticletitle{A Survey on Deep Learning for Software Engineering}.
\newblock \bibinfo{journal}{\emph{{ACM} Comput. Surv.}} \bibinfo{volume}{54}, \bibinfo{number}{10s} (\bibinfo{year}{2022}), \bibinfo{pages}{206:1--206:73}.
\newblock
\urldef\tempurl%
\url{https://doi.org/10.1145/3505243}
\showURL{%
\tempurl}


\bibitem[Yang et~al\mbox{.}(2023)]%
        {DBLP:journals/iacr/YangWCCY23}
\bibfield{author}{\bibinfo{person}{Zhenkun Yang}, \bibinfo{person}{Wen Wang}, \bibinfo{person}{Jeremy Casas}, \bibinfo{person}{Pasquale Cocchini}, {and} \bibinfo{person}{Jin Yang}.} \bibinfo{year}{2023}\natexlab{}.
\newblock \showarticletitle{Towards {A} Correct-by-Construction {FHE} Model}.
\newblock \bibinfo{journal}{\emph{{IACR} Cryptol. ePrint Arch.}} (\bibinfo{year}{2023}), \bibinfo{pages}{281}.
\newblock
\urldef\tempurl%
\url{https://eprint.iacr.org/2023/281}
\showURL{%
\tempurl}


\bibitem[Yasunaga and Liang(2020)]%
        {DBLP:conf/icml/YasunagaL20}
\bibfield{author}{\bibinfo{person}{Michihiro Yasunaga} {and} \bibinfo{person}{Percy Liang}.} \bibinfo{year}{2020}\natexlab{}.
\newblock \showarticletitle{Graph-based, Self-Supervised Program Repair from Diagnostic Feedback}. In \bibinfo{booktitle}{\emph{Proceedings of the 37th International Conference on Machine Learning, {ICML} 2020, 13-18 July 2020, Virtual Event}} \emph{(\bibinfo{series}{Proceedings of Machine Learning Research}, Vol.~\bibinfo{volume}{119})}. \bibinfo{publisher}{{PMLR}}, \bibinfo{pages}{10799--10808}.
\newblock
\urldef\tempurl%
\url{http://proceedings.mlr.press/v119/yasunaga20a.html}
\showURL{%
\tempurl}


\bibitem[Zhou et~al\mbox{.}(2022)]%
        {DBLP:journals/corr/abs-2211-09066}
\bibfield{author}{\bibinfo{person}{Hattie Zhou}, \bibinfo{person}{Azade Nova}, \bibinfo{person}{Hugo Larochelle}, \bibinfo{person}{Aaron~C. Courville}, \bibinfo{person}{Behnam Neyshabur}, {and} \bibinfo{person}{Hanie Sedghi}.} \bibinfo{year}{2022}\natexlab{}.
\newblock \showarticletitle{Teaching Algorithmic Reasoning via In-context Learning}.
\newblock \bibinfo{journal}{\emph{CoRR}}  \bibinfo{volume}{abs/2211.09066} (\bibinfo{year}{2022}).
\newblock
\urldef\tempurl%
\url{https://doi.org/10.48550/arXiv.2211.09066}
\showURL{%
\tempurl}


\end{thebibliography}


\end{document}